\documentclass[twocolumn, 10pt, pre, aps, showpacs, reprint, amsmath, amssymb, superscriptaddress, nofootinbib]{revtex4-1}

\usepackage[pdftex]{graphicx} 
\usepackage{subfigure}
\usepackage{bbm}

\begin{document}

\newcommand{\kt}{k_\parallel}
\newcommand{\lzt}{q_z}
\newcommand{\atled}{\bm{\nabla}}
\newcommand{\dx}{\frac{\partial}{\partial_x}}
\newcommand{\dy}{\frac{\partial}{\partial_y}}
\newcommand{\dz}{\frac{\partial}{\partial_z}}
\newcommand{\dt}{\frac{\partial}{\partial_t}}
\newcommand{\sqrdt}{\frac{\partial^2}{\partial_t^2}}
\newcommand{\pbyp}[2]{\frac{\partial #1}{\partial #2}}
\newcommand{\dbyd}[2]{\frac{d #1}{d #2}}
\newcommand{\ex}{\vec{e}_x}
\newcommand{\ey}{\vec{e}_y}
\newcommand{\ez}{\vec{e}_z}
\newcommand{\besselj}[2]{\mathrm{J}_{#1}(#2)}
\newcommand{\besseljp}[2]{\mathrm{J'}_{#1}(#2)}
\newcommand{\besseljs}[1]{\mathrm{J}_{#1}}
\newcommand{\besseljsp}[1]{\mathrm{J}_{#1}'}
\newcommand{\besseljspp}[1]{\mathrm{J}_{#1}''}
\newcommand{\hankel}[3]{\mathrm{H}_{#1}^{(#2)}(#3)}
\newcommand{\hankelp}[3]{\mathrm{H'}_{#1}^{(#2)}(#3)}
\newcommand{\hankels}[2]{\mathrm{H}_{#1}^{(#2)}}
\newcommand{\hankelsp}[2]{\mathrm{H}_{#1}'^{(#2)}}
\newcommand{\hankelspp}[2]{\mathrm{H}_{#1}''^{(#2)}}
\newcommand{\laplace}{\Delta}
\newcommand{\neff}{n_{\mathrm{eff}}}
\newcommand{\fexp}{f_{\mathrm{expt}}}
\newcommand{\ftheo}{f_{\mathrm{calc}}}
\newcommand{\ffit}{f_\mathrm{fit}}
\newcommand{\nexp}{\tilde{n}}
\newcommand{\Gtheo}{\Gamma_{\mathrm{calc}}}
\newcommand{\Gexp}{\Gamma_{\mathrm{expt}}}
\newcommand{\Grad}{\Gamma_{\mathrm{rad}}}
\newcommand{\Gabs}{\Gamma_{\mathrm{abs}}}
\newcommand{\Gant}{\Gamma_{\mathrm{ant}}}
\newcommand{\rhof}{\rho_{\mathrm{fluc}}}
\newcommand{\rhofscl}{\rhof^{\mathrm{scl}}}
\newcommand{\rhofSS}{\rhof^{(\mathrm{ss})}}
\newcommand{\rhofnr}[1]{\rho_{#1 n_r}}
\newcommand{\rhow}{\rho_{\mathrm{Weyl}}}
\newcommand{\Nweyl}{N_{\mathrm{Weyl}}}
\newcommand{\rhot}{\hat{\rho}}
\newcommand{\rhotscl}{\rhot_{\mathrm{scl}}}
\newcommand{\rhotSS}{\rhot^{(\mathrm{ss})}}
\newcommand{\rhotnr}[1]{\tilde{\rho}_{#1 n_r}}
\newcommand{\fmin}{f_{\mathrm{min}}}
\newcommand{\kmin}{k_{\mathrm{min}}}
\newcommand{\fmax}{f_{\mathrm{max}}}
\newcommand{\kmax}{k_{\mathrm{max}}}
\newcommand{\fcrit}{f_{\mathrm{co}}}
\newcommand{\kcrit}{k_{\mathrm{crit}}}
\newcommand{\kfilt}{k_\mathrm{filt}}
\newcommand{\po}{\mathrm{po}}
\newcommand{\lpo}{\ell_\po}
\newcommand{\lpeak}{\ell_\mathrm{peak}}
\newcommand{\lpeakscl}{\lpeak^\mathrm{scl}}
\newcommand{\lpeakexp}{\lpeak^\mathrm{expt}}
\newcommand{\lmax}{\ell_{\mathrm{max}}}
\newcommand{\Aexp}{\mathcal{A}_\mathrm{expt}}
\newcommand{\Ascl}{\mathcal{A}_\mathrm{scl}}
\newcommand{\alphacrit}{\alpha_{\mathrm{crit}}}
\newcommand{\alphainc}{\alpha_\mathrm{inc}}
\newcommand{\chico}{\chi_{\mathrm{co}}}
\newcommand{\Psiexp}{\Psi_\mathrm{expt}}
\newcommand{\Psit}{\tilde{\Psi}}
\newcommand{\Psitexp}{\Psit_\mathrm{expt}}
\newcommand{\Psimod}{\Psi_\mathrm{mod}}
\newcommand{\Psisup}{\Psi_\mathrm{sup}}
\newcommand{\reffig}[1]{\mbox{Fig.~\ref{#1}}}
\newcommand{\subreffig}[1]{\mbox{Fig. \subref{#1}}}
\newcommand{\refeq}[1]{\mbox{Eq.~(\ref{#1})}}
\newcommand{\refsec}[1]{\mbox{Sec.~\ref{#1}}}
\newcommand{\reftab}[1]{\mbox{Table \ref{#1}}}
\newcommand{\etal}{\textit{et al.\ }}
\newcommand{\dA}{A}
\newcommand{\dB}{B}
\newcommand{\FSR}{\mathrm{FSR}}
\newcommand{\depth}{l}
\newcommand{\afoam}{d}
\newcommand{\shv}{s_{xy}}
\newcommand{\Exy}{E_{m_x, m_y}}
\newcommand{\gxy}{g_{m_x, m_y}}
\newcommand{\Pone}{\mathcal{P}_1}
\newcommand{\Ptwo}{\mathcal{P}_2}
\renewcommand{\Re}[1]{\mathrm{Re}\left(#1\right)}
\renewcommand{\Im}[1]{\mathrm{Im}\left(#1\right)}

\hyphenation{re-so-nan-ce re-so-nan-ces ex-ci-ta-tion z-ex-ci-ta-tion di-elec-tric ap-pro-xi-ma-tion ra-dia-tion Me-cha-nics quan-tum pro-posed Con-cepts pro-duct Reh-feld ob-ser-va-ble Se-ve-ral rea-so-nable Ap-pa-rent-ly re-pe-ti-tions re-la-tive quan-tum su-per-con-duc-ting ap-pro-xi-mate cri-ti-cal mea-su-red}

\title{Dielectric square resonator investigated with microwave experiments}

\author{S. Bittner}
\affiliation{Laboratoire de Photonique Quantique et Mol{\'e}culaire, CNRS UMR 8537, Institut d'Alembert FR 3242, {\'E}cole Normale Sup{\'e}rieure de Cachan, F-94235 Cachan, France}
\author{E. Bogomolny}
\affiliation{Universit{\'e} Paris-Sud, LPTMS, CNRS UMR 8626, Orsay, F-91405, France}
\author{B. Dietz}
\email{dietz@ikp.tu-darmstadt.de}
\author{M. Miski-Oglu}
\affiliation{Institut f\"ur Kernphysik, Technische Universit\"at Darmstadt, D-64289 Darmstadt, Germany}
\author{A. Richter}
\email{richter@ikp.tu-darmstadt.de}
\affiliation{Institut f\"ur Kernphysik, Technische Universit\"at Darmstadt, D-64289 Darmstadt, Germany}

\date{\today}

\begin{abstract}
We present a detailed experimental study of the symmetry properties  and the momentum space representation of the field distributions of a  dielectric square resonator as well as the comparison with a  semiclassical model. The experiments have been performed with a flat ceramic microwave resonator. Both the resonance spectra and the field distributions were measured. The momentum space representations of the latter evidenced that the resonant states are each related to a specific classical torus, leading to the regular structure of the spectrum. Furthermore, they allow for a precise determination of the refractive index. Measurements with different arrangements of the emitting and the receiving antennas were performed and their influence on the symmetry properties of the field distributions was investigated in detail, showing that resonances with specific symmetries can be selected purposefully. In addition, the length spectrum deduced from the measured resonance spectra and the trace formula for the dielectric square resonator are discussed in the framework of the semiclassical model.
\end{abstract}

\pacs{05.45.Mt, 03.65.Sq, 03.65.Ge}

\maketitle

\section{Introduction}
The interest in the resonance properties of dielectric resonators stems from their scattering properties \cite{Mie1908, Nelson1976}, their use as compact resonators or filters in electronic rf circuits \cite{Kajfez1998} and applications like lasers or sensors of microscopic dimensions in the infrared to optical frequency regime \cite{Vahala2004, Matsko2009}. Flat microlasers with a large diversity of shapes have been investigated in order to understand the connections between their properties (e.g., emission directionality and threshold) and their shape \cite{Wiersig2011b}. Even in two dimensions, the Helmholtz equation describing the passive resonators can be solved analytically only for a few cases of rotationally symmetric structures \cite{Mie1908, Hentschel2002b}. Hence, generally, numerical methods like the boundary element method \cite{Wiersig2003a, Smotrova2013}, the finite difference time domain method \cite{Vuckovic1999}, or perturbation theory \cite{Dubertrand2008, Ge2013, Ge2013a} are indispensable for their theoretical analysis. Cavities with sizes much larger than the wavelength provide an exception since they can be considered as photon billiards, that is, their properties can be related to the classical ray dynamics in the framework of semiclassical approximations \cite{Schwefel2003}. Such models, though approximate, generally allow for a better understanding of the resonance spectra and field distributions than solely numerical methods and can be extended to size-to-wavelength regimes where numerical computations are no longer feasible.

A phenomenon of particular interest are resonant states exhibiting exceptionally clear structures associated with classical trajectories like periodic orbits (POs). Examples are wave functions that are enhanced around unstable POs, called scars \cite{Heller1984, Gmachl2002, Harayama2003}, or Gaussian modes related to stable POs \cite{Tureci2002a, Shinohara2011}. Furthermore, resonant states concentrated on trajectories with a specific (angular) momentum have been observed for oval and square dielectric resonators and circular resonators with rough boundaries \cite{Noeckel2001, Bittner2013b, Fang2005}. It was shown in Refs.~\cite{Bogomolny2004a, Bogomolny2006} that closed polygonal resonators with angles $\pi m_j / n_j$, where $m_j,\, n_j$ are coprime integers and $m_j \ne 1$ exhibit a pronounced scarring behavior of the wave functions which can be related to families of POs with parallel trajectories and attributed to the strong diffraction at the corners. Similar modes exist in polygonal dielectric resonators \cite{Lebental2007, Song2013a}. Resonators with polygonal shape are of importance, e.g., for filter applications \cite{Li2006, Marchena2008, Yan2013} or because crystalline materials favor the fabrication of polygonal cavities \cite{Vietze1998, Chang2000, Nobis2004, Wang2006}. Cavities with regular polygonal shape have been intensively investigated, and a large number of in many cases similar ray-based models for the resonant modes of equilateral triangles \cite{Chang2000, Wysin2006, Yang2009a, Yang2009b}, squares \cite{Marcatili1969, Poon2001, Fong2003, Guo2003, Moon2003, Chern2004, Li2006, Lebental2007, Huang2008, Yang2009a, Che2010a, Bittner2010, Bittner2011a}, and hexagons \cite{Wiersig2003, Nobis2004, Wang2006, Liu2008} have been proposed.

In this article we investigate experimentally the structure of the modes of a dielectric square resonator with a microwave experiment. The cavity is made from a low-loss ceramic material and is large compared to the wavelengths of the microwaves coupled into it. One advantage of microwave resonators is that the field distributions inside the resonator can be measured directly and the results of these studies can be applied to microcavities since these have a similar ratio between cavity size and wavelength. Furthermore, the dimensions of a macroscopic resonator can be measured with high precision and geometric imperfections that would break the symmetry properties can be excluded.

In the experiments with a square resonator presented here, all observed modes were related to specific classical tori, i.e., families of orbits that are defined by their common angle of incidence \cite{Bittner2013b}. This is surprising because generally only a few modes of a cavity show a structure which can be related to classical trajectories \cite{Bogomolny1988}, like, e.g., the scars observed in the stadium billiard \cite{Heller1984, Mcdonald1988}. We demonstrate that they can be described by a semiclassical model proposed in Ref.~\cite{Bittner2013b} in a unified way and present a detailed comparison of different aspects of the model with the experimental data.

The article is organized as follows. In \refsec{sec:expSetup} the experimental setup and in \refsec{sec:resTheo} the general theoretical framework for flat dielectric resonators as well as the ray-based model from Ref.~\cite{Bittner2013b} are introduced. The measured frequency spectra are discussed in \refsec{sec:freqSpec}. In \refsec{sec:fieldDistr}, the measurement of the field distributions and their analysis is detailed. Section \ref{sec:expVsMod} presents an overview over the experimental data and an in-depth comparison with the ray-based model. The length spectrum and the trace formula for the dielectric square resonator are discussed in \refsec{sec:lspect}, and \refsec{sec:concl} ends with some concluding remarks.

\section{Experimental Setup} \label{sec:expSetup}

\begin{figure}[tb]
\begin{center}
\subfigure[]{
	\includegraphics[width = 7.0 cm]{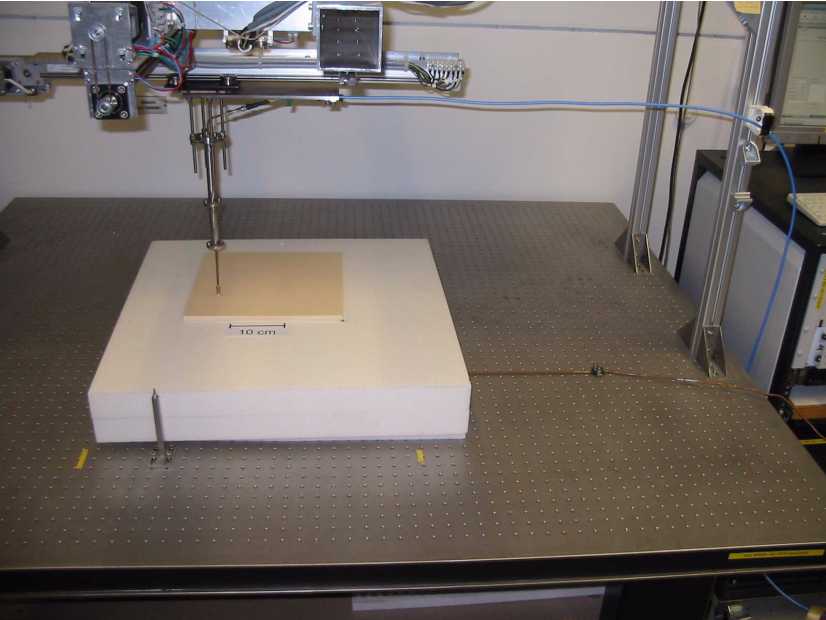}
	\label{sfig:setupPhoto}
}
\subfigure[]{
	\includegraphics[width = 7.0 cm]{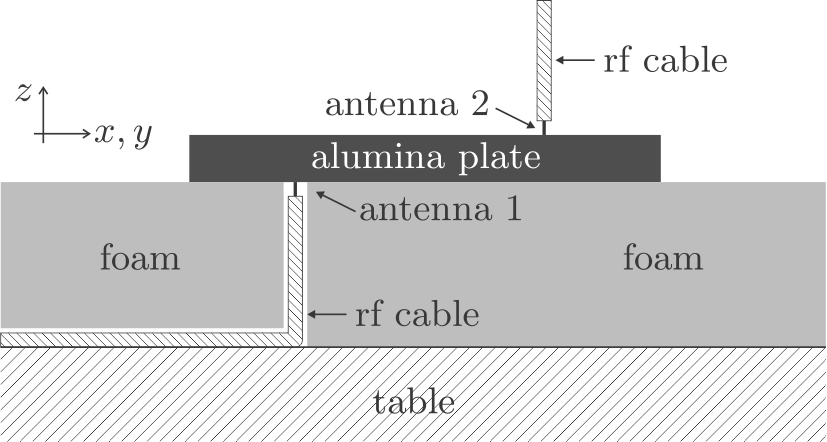}
	\label{sfig:setupSketch}
}
\subfigure[]{
	\includegraphics[width = 5 cm]{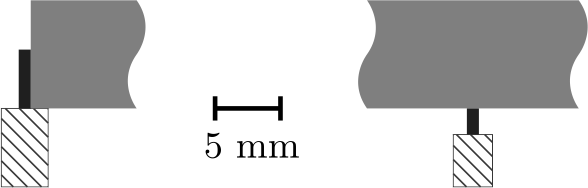}
	\label{sfig:setupAntennas}
}
\end{center}
\caption{(Color online) Experimental setup. \subref{sfig:setupPhoto} Photograph of the experimental setup. \subref{sfig:setupSketch} Sketch (not to scale) of the setup (reprinted from Ref.~\cite{Bittner2013b}). \subref{sfig:setupAntennas} Drawings of the two antenna configurations used, i.e., the antenna was either placed at the side wall of (left) or below (right) the ceramic plate.}
\label{fig:setup}
\end{figure}

Figure \ref{fig:setup} shows a photograph and a sketch of the experimental setup. A ceramic plate made of alumina (Deranox 995 by Morgan Advanced Ceramics with 99.5\% Al$_2$O$_3$ content) was used as microwave resonator. The plate was manufactured precisely to be square and to have sharp corners and edges (the deviation of the angles from $90^\circ$ is less than $0.1^\circ$). It had a side length of $a = (297.30 \pm 0.05)$ mm and a thickness of $b = (8.27 \pm 0.01)$ mm. In the frequency range of interest its refractive index  was determined as $n_1 = 3.10$ (see \refsec{sec:fieldDistr}). The plate sat atop a $\afoam = 120.0$ mm thick layer of foam (Rohacell 31IG by Evonik Industries \cite{Rohacell}) with refractive index $n_2 = 1.02$ and low absorption such that a free-floating resonator was effectively simulated.

Two wire antennas protruding from coaxial cables were coupled to the resonator and aligned perpendicularly to its plane. A vectorial network analyzer (VNA, model PNA N5230A by Agilent Technologies) was used to measure the complex transmission amplitude $S_{21}$ from antenna $1$ to antenna $2$. The excitation antenna $1$ was put at different positions either next to a side wall of the resonator or below it [see left and right part of \reffig{sfig:setupAntennas}, respectively] in order to excite resonances of specific symmetry classes (see \refsec{sec:freqSpec}). The receiving antenna $2$ could be moved around by a computer controlled positioning unit, allowing to map out the field distributions (see \refsec{sec:fieldDistr}). It had a configuration as shown in the right panel of \reffig{sfig:setupAntennas}, though coming from above instead of from below, with an additional Teflon hat in order to reduce the friction with the ceramic plate. Before the measured frequency spectra are discussed in \refsec{sec:freqSpec}, we will briefly review the general modeling of flat dielectric resonators in the next section.

\section{Modeling of the resonator} \label{sec:resTheo}

\subsection{Effective refractive index approximation} \label{ssec:neffApp}
Electromagnetic resonators are described by the well-known vectorial Helmholtz equation with appropriate boundary conditions which is difficult to solve for general three-dimensional (3D) resonators. The dielectric resonator considered in the present article has a cylindrical geometry and is flat, that is, it has a thickness $b$ of the order of the wavelength $\lambda$ or smaller and a transverse extension much larger than the wavelength. Such resonators can be approximated as two-dimensional (2D) systems \cite{Vassallo1991, Lebental2007, Bittner2009} by introducing an effective refractive index $\neff$. In the framework of this approximation, the resonator is considered as an infinite slab waveguide where the phase velocity with respect to the $x y$ plane [cf.~\reffig{sfig:setupSketch}] is $c / \neff$ with $c$ the speed of light in vacuum and the modes in the resonator can be separated into modes with transverse magnetic (TM) and with transverse electric (TE) polarization having a magnetic field $\vec{B}$ and, respectively, an electric field $\vec{E}$ parallel to the plane of the resonator. The ansatz for the field component $E_z$ ($B_z$) inside the resonator for modes with TM (TE) polarization is
\begin{equation} \left. \begin{array}{c} E_z \\ B_z \end{array} \right\} = \Psi(x, y) (A_1 e^{i k_z z} + A_2 e^{-i k_z z}) \end{equation}
where $A_{1,2}$ are constants and $\Psi$ is called the wave function (WF) in the following. The $z$ component $k_z$ of the wave vector is related to $\neff$ via $k_z = k (n_1^2 - \neff^2)^{1/2}$, where $n_1$ is the refractive index of the resonator material and $k$ is the wave number that is related to the frequency via $2 \pi f = c k$. We only consider modes guided by total internal reflection (TIR) here. Their fields decay exponentially with $\exp\{ -|z| / (2 \, l_{2, 3}) \}$ above and below the resonator where the decay lengths are 
\begin{equation} \label{eq:decayLength} l_{2, 3} = \left( 2 k \sqrt{\neff^2 - n_{2, 3}^2} \right)^{-1} \, . \end{equation}
Here $n_{2, 3}$ are the refractive indices of the material below and above the resonator, respectively. Matching of the ansatz for the fields inside and outside the resonator yields the quantization condition for the effective refractive index \cite{Lebental2007, Bittner2009, Bittner2012a}. The vectorial Helmholtz equation for the resonator thus reduces to the 2D scalar Helmholtz equation,
\begin{equation} \label{eq:helmholtzScalar} \left( \Delta + \neff^2 k^2 \right) \Psi(x, y) = 0 \end{equation}
for the WF $\Psi(x, y)$ associated with $E_z$ ($B_z$) for TM (TE) modes, where $\kt = \neff k$ is the component of the wave vector parallel to the plane of the resonator. The boundary conditions are that $\Psi$ is continuous at the boundary of the resonator in the case of both polarizations and that $\partial \Psi / \partial n$ is continuous for TM modes while $\mu (\partial \Psi / \partial n)$ is continuous for TE modes, with $\vec{n}$ the normal vector of the boundary, $\mu = \neff^{-2}$ inside the resonator, and $\mu = 1$ outside. It should be noted, however, that the $\neff$ approach, even though it turned out to be most suitable for the present considerations, is an approximation which, e.g., results in small deviations between the predicted and the actual resonance frequencies (see \refsec{ssec:dataReview}). Furthermore, the behavior of the fields close to the boundary is not well understood since there the approximation resulting from the assumption that the resonator can be regarded as an infinite slab waveguide no longer applies. The boundary conditions for the field components $E_z$ and $B_z$ are actually coupled at the side walls of the resonator \cite{Schwefel2005} so there the separation into TM and TE modes is only approximate. 

For the resonator configuration depicted in \reffig{fig:setup}, the ansatz for the fields below the resonator needs to be slightly modified due to the presence of the metal table. This yields the quantization condition \cite{Bittner2012a}
\begin{equation} \label{eq:neffQuant} k b = \left\{ \delta_{12} + \delta_{13} + \zeta \pi \right\} / \sqrt{n_1^2 - \neff^2} \, . \end{equation}
where $\zeta = 0, \, 1, \, 2, \dots$ is the $z$ quantum number counting the number of field nodes in the $z$ direction. The two phases are
\begin{equation} \delta_{13} = \arctan\left\{ \nu_{13} \sqrt{\frac{\neff^2 - n_3^2}{n_1^2 - \neff^2}} \right\} \end{equation}
and
\begin{equation} \delta_{12} = \arctan\left\{ \nu_{12} \sqrt{\frac{\neff^2 - n_2^2}{n_1^2 - \neff^2}} \, h\left( \frac{d}{2 l_2} \right) \right\} \end{equation}
with $\nu_{1j} = n_1^2 / n_j^2$ for TM and $\nu_{1j} = 1$ for TE modes, respectively. Here $n_2 = 1.02$ is the refractive index of the foam and $n_3 = 1$ that of air. The function $h(x)$ in $\delta_{12}$ is $h(x) = \tanh(x)$ for TM and $h(x) = \coth(x)$ for TE modes, respectively, and accounts for the boundary conditions at the metal table. For a sufficiently short decay length $l_2$, $d \gg 2 l_2$, the influence of the optical table becomes negligible since $h[d / (2 l_2)] \approx 1$. 

In the experiments described in the present article, both TM and TE modes with different $z$ excitations, denoted as TM$_\zeta$ and TE$_\zeta$ in the following, were observed, but the investigations concentrate on the TM$_0$ modes for which the best data is available. Furthermore, the effective refractive index depends strongly on the frequency. For the TM$_0$ modes, $\neff$ takes values between $1.5$ and $2.5$ in the frequency range of interest. In fact, it was determined experimentally from the measured field distributions, and the theoretical curve for $\neff(f)$ was fitted to the experimental data to obtain the refractive index $n_1$ of the alumina more precisely than provided by the manufacturer (see \refsec{sec:fieldDistr}).

\subsection{Ray-based model for the dielectric square resonator} \label{ssec:rayModel}

\begin{table*}[tb]
\caption{Symmetry classes, quantum numbers and model WFs. The first column denotes the symmetry with respect to the diagonals, the second column is the symmetry with respect to both the horizontal and vertical axis, the third column is the parity of $m_x + m_y$, the fourth column the parity of $m_x$ and $m_y$ [which is the same for $(++)$ and $(--)$ modes but different for $(+-)$ and $(-+)$ modes], the fifth column is the Mulliken symbol, and the sixth column gives the corresponding model WF (adapted from Ref.~\cite{Bittner2013b}).}
\label{tab:dlmWFs}
\vspace{3 mm}
\begin{center}
\begin{tabular}{c|c|c|c|c|l}
\hline
\hline
Diagonal & Horizontal/vertical & Parity of & Parity of & Mulliken & Model wave function \\
symmetry & symmetry & $m_x + m_y$ & $m_x$, $m_y$ & symbol & \\
\hline
$(++)$ & $+$ & Even & Even & $A_1$ & $\Psimod(x, y) = \cos(k_x x) \cos(k_y y) + \cos(k_y x) \cos(k_x y)$ \\
$(++)$ & $-$ & Even & Odd & $B_1$ & $\Psimod(x, y) = \sin(k_x x) \sin(k_y y) + \sin(k_y x) \sin(k_x y)$ \\
$(--)$ & $+$ & Even & Even & $B_2$ & $\Psimod(x, y) = \cos(k_x x) \cos(k_y y) - \cos(k_y x) \cos(k_x y)$ \\
$(--)$ & $-$ & Even & Odd & $A_2$ & $\Psimod(x, y) = \sin(k_x x) \sin(k_y y) - \sin(k_y x) \sin(k_x y)$ \\
$(+-)$ & None & Odd & & $E$ & $\Psimod(x, y) = \sin(k_x x) \cos(k_y y) + \cos(k_y x) \sin(k_x y)$ \\
$(-+)$ & None & Odd & & $E$ & $\Psimod(x, y) = \sin(k_x x) \cos(k_y y) - \cos(k_y x) \sin(k_x y)$ \\
\hline
\hline
\end{tabular}
\end{center}
\end{table*}

In order to form a resonant state, a wave traveling along a trajectory must be phase-matched after one round trip through the resonator. This leads to the approximate quantization condition \cite{Bittner2013b}
\begin{equation} \label{eq:quantCond} \begin{array}{rcl} \exp\{ 2 i k_x a \} r^2(\alpha_x) & = & 1 \\ \exp\{ 2 i k_y a \} r^2(\alpha_y) & = & 1 \, . \end{array} \end{equation}
The momentum vector components $k_{x, y}$ of the wave are related to the wave number via $k = (k_k^2 + k_y^2)^{1/2} / \neff$, and $\ftheo = c k / (2 \pi)$ is the corresponding resonance frequency. In the case of TM modes ($s$ polarization), the corresponding Fresnel coefficients,
\begin{equation} r(\alpha) = \frac{\neff \cos(\alpha) - \sqrt{1 - \neff^2 \sin^2(\alpha)}}{\neff \cos(\alpha) + \sqrt{1 - \neff^2 \sin^2(\alpha)}} \, , \end{equation}
account for the (partial or total) reflections at the cavity boundaries.

We use the Fresnel reflection coefficients for an infinite interface because those for a finite interface are nontrivial. Nevertheless, the agreement between the model and the experiment turned out to be good. The angles of incidence with respect to normal vectors on the boundaries perpendicular to the $x$ axis (respectively, the $y$) axis are 
\begin{equation}\alpha_{x, y} = \arctan[\Re{k_{y, x}} / \Re{k_{x, y}}] \, . \end{equation}
It should be noted that the dependence of the effective refractive index on the frequency (respectively, the wave number) must be taken into account when solving the quantization condition, \refeq{eq:quantCond}. Its solutions can be written as 
\begin{equation} \label{eq:quantCondSol} \begin{array}{rcl} k_x & = & \{ \pi m_x + i \ln[r(\alpha_x)] \} / a \\ k_y & = & \{ \pi m_y + i \ln[r(\alpha_y)] \} / a \, . \end{array} \end{equation}
Accordingly each mode can be labeled by its symmetry class and the $x$ and $y$ quantum numbers $m_{x, y} = 0, 1, 2, \dots$, where the case $(m_x, m_y) = (0, 0)$ must be excluded. This was confirmed experimentally in Ref.~\cite{Bittner2013b}, i.e., the resonant modes of the dielectric square resonator are associated with specific classical tori that consist of nonclosed trajectories having the same angles of incidence. Note that even though only the TM modes are discussed in this paper, the model should also apply for TE modes. 

The model WFs $\Psimod(x, y)$ are composed of a superposition of eight plane waves with wave vectors $(\pm k_x, \pm k_y)$ and $(\pm k_y, \pm k_x)$ determined by the set of classical trajectories that they are related to. The relative phases of the different plane wave components are fixed by the symmetry of the WF. Since the square resonator belongs to the point group $C_{4v}$, its modes belong to six different symmetry classes \cite{McIsaac1975, Guo2003, Yang2007}. An overview over the different symmetry classes (i.e., irreducible representations of the point group $C_{4v}$) and the corresponding model WFs is given in \reftab{tab:dlmWFs}. The first column gives the reflection symmetry with respect to the diagonals of the square, $(s_1 s_2)$ with $s_{1, 2} \in \{+, -\}$, where $s_1 = +1$ ($s_2 = +1$) when the WF of the mode is symmetric, and $s_1 = -1$ ($s_2 = -1$) when it is antisymmetric with respect to the diagonal $x = y$ ($x = -y$). The second column denotes the mirror symmetry $s_x$ ($s_y$) with respect to the $x = 0$ ($y = 0$) axis, where $s_x = s_y = +1$ ($s_x = s_y = -1$) for symmetric (antisymmetric) WFs. The third and fourth columns contain the conditions for the parity of the quantum numbers for the different symmetry classes. Each mode can thus be labeled unambiguously by $(m_x, m_y, s_1 s_2)$. In the fifth column the Mulliken symbols for the different representations of the group $C_{4v}$ are given \cite{Yang2007}. The sixth column contains the model WFs $\Psimod(x, y)$. The notations $(m_x, m_y, s_1 s_2)$ and $(m_y, m_x, s_1 s_2)$ refer to the same mode, where we us $m_x \leq m_y$. 

The modes of the $E$ representation [i.e., with $(+-)$ and $(-+)$ symmetry] are degenerate due to their symmetry \cite{McIsaac1975}. Therefore, for the $E$ representations the assignment of the model WFs is not unambiguous, and other WFs belonging to these representations can be constructed as superpositions of the model WFs given in \reftab{tab:dlmWFs}. This includes, in particular, WFs that are symmetric with respect to the horizontal axis and antisymmetric with respect to the vertical one (or vice versa) but have no well-defined symmetry with respect to the diagonals. Actually the shape of the WFs of the $E$ representation depends, e.g., on the manner of excitation or on small perturbations (see \refsec{ssec:symProp}). The model also predicts that the modes $(m_x, m_y, --)$ and $(m_x, m_y, ++)$ are degenerate. In practice, however, there is a small difference between their resonance frequencies that stems from the fact that the $(--)$ modes have a vanishing WF at the corners, whereas that of the $(++)$ modes is nonvanishing. Furthermore, modes with $m_x = m_y$ always have $(++)$ symmetry.

We define the overlap between two (normalized) WFs $\Psi_{1, 2}$ as the modulus squared of the overlap coefficient,
\begin{equation} C_{12} = \left< \Psi_1 \middle| \Psi_2 \right> = \int \limits_{-a/2}^{a/2} dx \int \limits_{-a/2}^{a/2} dy \, \Psi_1^*(x, y) \Psi_2(x, y) \, . \end{equation}
It should be noted that the different model modes are not exactly orthogonal; however, their mutual overlaps were always smaller than $1 \%$ in the cases considered here. The family of trajectories to which a mode is related, and hence the mode itself, can be characterized by the angle of incidence,
\begin{equation} \alphainc = \min \{ \alpha_x, \alpha_y \} \approx \arctan(m_x / m_y) \, , \end{equation}
where $0^\circ \leq \alphainc \leq 45^\circ$. Modes with $m_x \approx m_y$ are therefore associated to trajectories close to the family of the diamond periodic orbit, i.e., POs that are reflected once at each side of the resonator with an angle of incidence of $45^\circ$. This type of modes is the most commonly observed one \cite{Guo2003a, Moon2003, Yang2009a, Bittner2011a}, in particular for systems with a relatively low refractive index \cite{Pan2003, Chern2004, Lebental2007, Bittner2010}. Models based on the diamond orbit can be derived on the basis of that introduced in the present article. This will be further discussed in \refsec{ssec:modID} (see also Ref.~\cite{Guo2003}).

The solutions $k_{x, y}$ of \refeq{eq:quantCond} are in general complex, i.e., the modes have a finite lifetime due to refractive losses. The associated quality factors are
\begin{equation} Q = - \Re{\ftheo} / [2 \, \Im{\ftheo}] \, . \end{equation}
If $\alphainc > \alphacrit = \arcsin(1 / \neff)$, the trajectories are confined in the resonator by total internal reflection (TIR). Then the Fresnel coefficients $r(\alpha_{x, y})$ have unit modulus, and the terms $i \ln[r(\alpha_{x, y})]$ in \refeq{eq:quantCondSol} are purely real and signify the phase shift at the reflection. Hence the model predicts for the associated modes purely real momentum vectors and wave functions, that is, infinite lifetimes. In reality, however, this is not the case due to diffractive losses that are not taken into account by the  model. An extension that includes these will be published elsewhere  \cite{BittnerPrep}. Nonetheless, because refractive losses are absent for modes confined by TIR they have longer lifetimes and smaller imaginary parts of $k_{x, y}$ than those with $\alphainc < \alphacrit$. Actually, all modes that were observed experimentally belong to the set of confined modes (see~\refsec{ssec:dataReview}) and even though the model cannot correctly predict the imaginary parts of $k_{x, y}$, the model WFs agree well with the measured ones as will be shown in \refsec{ssec:modID}.

\section{Measured frequency spectra} \label{sec:freqSpec}

\begin{figure*}[tb]
\begin{center}
\includegraphics[width = 16 cm]{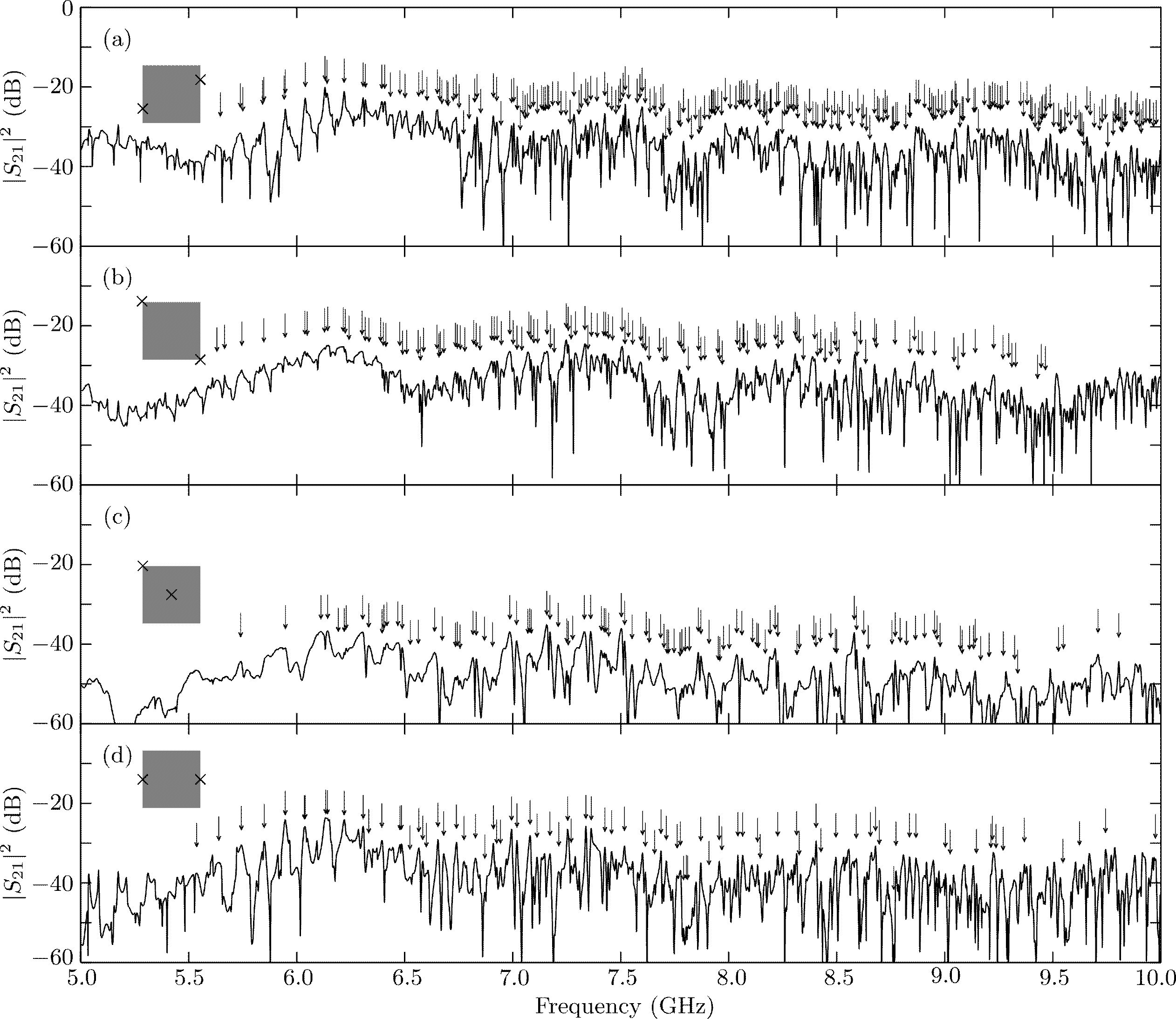}
\end{center}
\caption{Measured frequency spectra for different antenna combinations. The resonances identified as TM$_0$ modes are indicated by the arrows. The insets indicate the positions of the antennas at the side walls, respectively, below the cavity. (a) The excitation antenna was placed $a/4$ above the lower left corner while the receiving antenna was placed $a/4$ below the upper right corner. (b) The excitation antenna was positioned at the upper left and the receiving antenna at the lower right corner. (c) The excitation antenna was put beneath the center of the cavity and the receiving antenna at the upper left corner. (d) The excitation antenna was placed at the midpoint of the left and the receiving antenna at the midpoint of the right cavity edge.}
\label{fig:freqSpectra}
\end{figure*}

Four examples of frequency spectra measured with different positions of the antennas are shown in \reffig{fig:freqSpectra}. The positions of the antennas that were used are indicated in the insets as crosses. When the excitation antenna was placed at an edge or corner of the cavity, the antenna was coupled to the resonator as shown in the left panel of \reffig{sfig:setupAntennas}, otherwise as depicted in the right panel. The spectra display a large number of resonances with quality factors in the range of $Q = 500$--$2000$. The latter are bounded due to absorption in the ceramic material and coupling losses induced by the antennas. The resonance density grows with increasing frequency, and therefore more and more resonances are partially overlapping. The spectra also feature a background that varies slowly on the scale of $1$--$2$~GHz and results from direct transmission processes between the antennas. The polarization (TM or TE) of the resonant modes was determined with the perturbation technique detailed in Ref.~\cite{Bittner2012a}. Those that were identified as TM$_0$ modes are indicated by arrows. In the considered frequency range from $\fmin = 5.5$~GHz to $\fmax = 10.0$~GHz (where $\fmax$ corresponds to $k a = 62.3$) most modes were of the TM$_0$ type. Modes with higher $z$ excitation, i.e., TM$_{1, \, 2}$, were observed only above $\approx 8$~GHz. In addition, several TE modes were found, even though the vertical wire antennas couple preferentially to TM modes. That these antennas can also excite TE modes was already observed in Ref.~\cite{Bittner2009}.

A resonant state with TM polarization only can be excited if its electric field component $E_z$ is nonvanishing at the position of one of the antennas \cite{Stein1995}. Thus the positions of the antennas determine the symmetry class of the resonant modes that can be excited and observed. Accordingly, the antenna configurations (a)--(d) [corresponding to the insets in \reffig{fig:freqSpectra}(a)--\reffig{fig:freqSpectra}(d)] can couple only to modes of certain symmetry classes. This becomes noticeable in the total number of TM$_0$ modes that were observed in the corresponding spectra. Antenna configuration (b), for example, can only couple to modes with nonvanishing wave function along the diagonal $x = -y$, i.e., with $s_2 = +1$. These are the modes belonging to the $(++)$ and $(-+)$ symmetry classes ($A_1$, $B_1$, and $E$ representations). Configuration (c) only allows modes belonging to the $A_1$ representation. It corresponds to the most restricted case and thus leads to the sparsest spectrum of the four. Configuration (d) only couples to modes that are not antisymmetric with respect to the horizontal axis, that is, the $A_1$, $B_2$, and $E$ representations. The antennas of configuration (a) finally can couple to modes of all symmetry classes since they are not situated on any symmetry axis of the square resonator. Correspondingly, the spectrum shown in \reffig{fig:freqSpectra}(a) exhibits the largest resonance density of TM$_0$ modes of the four spectra.

As already mentioned in \refsec{ssec:rayModel}, the doubly degenerate modes of the $E$ representation can in general exhibit wave functions with various mirror symmetries depending on the excitation scheme. Antenna configuration (b) can only couple to modes with $s_2 = +1$, and the modes of the $E$ representation excited by it thus must be represented by the model WFs given in the last line of \reftab{tab:dlmWFs}. Configuration (d), on the other hand, selects WFs that are symmetric with respect to the horizontal axis and are therefore antisymmetric with respect to the vertical axis while having no defined symmetry with respect to the diagonal axes. These WFs correspond to specific superpositions of the model WFs given for the modes of the $E$ representation. The modes of that type excited by antenna configuration (a) have no defined symmetry with respect to any of the symmetry axes and therefore also correspond to superpositions of the model WFs. This will be further discussed in \refsec{ssec:symProp}. The next section describes the technique used for the measurement of the WFs.

\section{Measurement and analysis of field distributions} \label{sec:fieldDistr}

The field distributions of the resonant states inside the resonator are measured by the scanning antenna method. The position of one antenna, $1$, is kept fixed while the other one, $2$, is moved along the resonator surface. This technique exploits that at the resonance frequency $f_j$ of the resonant state $j$ the transmission amplitude $S_{21}$ between the vertical wire antennas $1$ and $2$ is related to the electric field distribution $E_z$ at the antenna positions $\vec{r}_{1, 2}$ by \cite{Stein1995}
\begin{equation} \label{eq:fieldMeas} S_{21}(f_j) \propto E_z(\vec{r}_2) \, E_z(\vec{r}_1) \, . \end{equation}
Hence $S_{21}(\vec{r}_2, f_j)$ is directly proportional to the electric field distribution $E_z(\vec{r}_2)$ and we may identify it with the wave function of the resonance $j$, $\Psiexp^{(j)}[\vec{r}_2 = (x, y)]$, for TM-polarized modes. Both the amplitude and phase of the signal transmitted from antenna $1$ to antenna $2$, and thus those of the complex WFs $\Psiexp$, are measured. 

It should be noted that the moving antenna induces a frequency shift and a broadening of the resonances depending on its position \cite{Stein1995, Tudorovskiy2008}. In practice, however, these effects are sufficiently small because the antennas only couple to the evanescent fields above and below the resonator so we can nonetheless consider $S_{21}(\vec{r}_2, f_j)$ as the measured WF in good approximation. The relation \refeq{eq:fieldMeas} between the transmission amplitude and the WF of a resonance breaks down, however, in the case of strongly overlapping or degenerate resonances \cite{Kuhl2000, Tudorovskiy2008, Tudorovskiy2011}. This applies especially to the case of the doubly degenerate modes of the $E$ representation. In this case the problem can be circumvented by placing the excitation antenna on a symmetry axis of the resonator, e.g., use antenna configurations (b) or (d), so only the mode of the degenerate pair that does not have a nodal line on this axis can be excited. 

As noted above, the wire antennas also couple to the TE-polarized resonant states. We suspect that the wire antennas also slightly couple to other components of the electric or magnetic field vectors. Since the details of the coupling mechanism are not understood for the TE modes; however, we cannot properly interpret the meaning of $S_{21}(\vec{r}_2, f_j)$ in those cases and thus discuss only the TM modes in the following. Furthermore, direct transmission processes between the antennas may contribute to the measured WFs.

\begin{figure}[tb]
\begin{center}
\subfigure[]{
	\includegraphics[width = 8.0 cm]{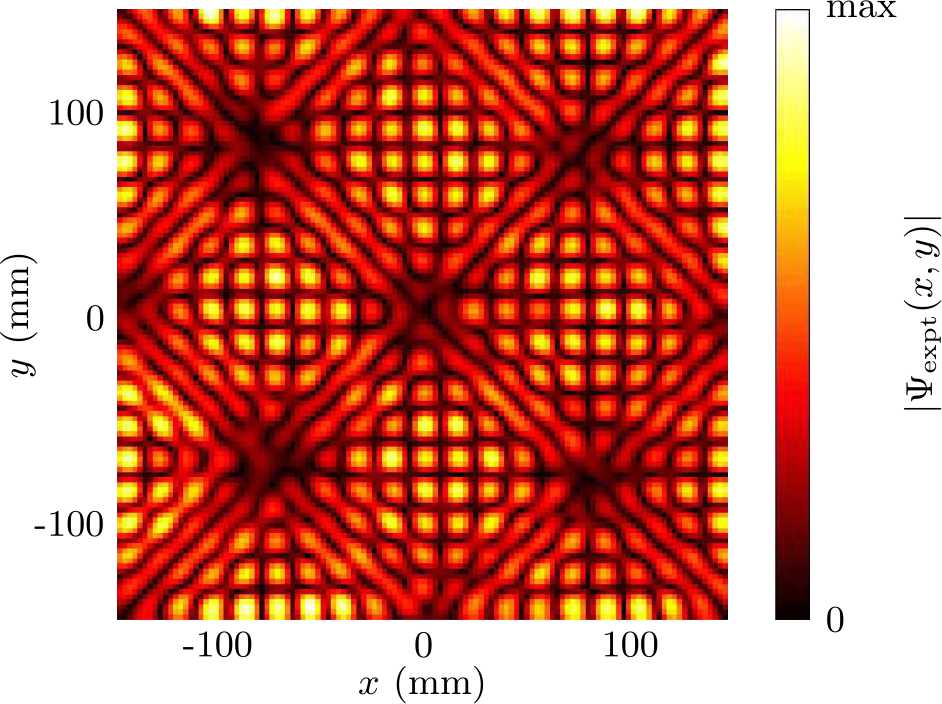}
	\label{sfig:WFgenCS} 
}
\subfigure[]{
	\includegraphics[width = 8.0 cm]{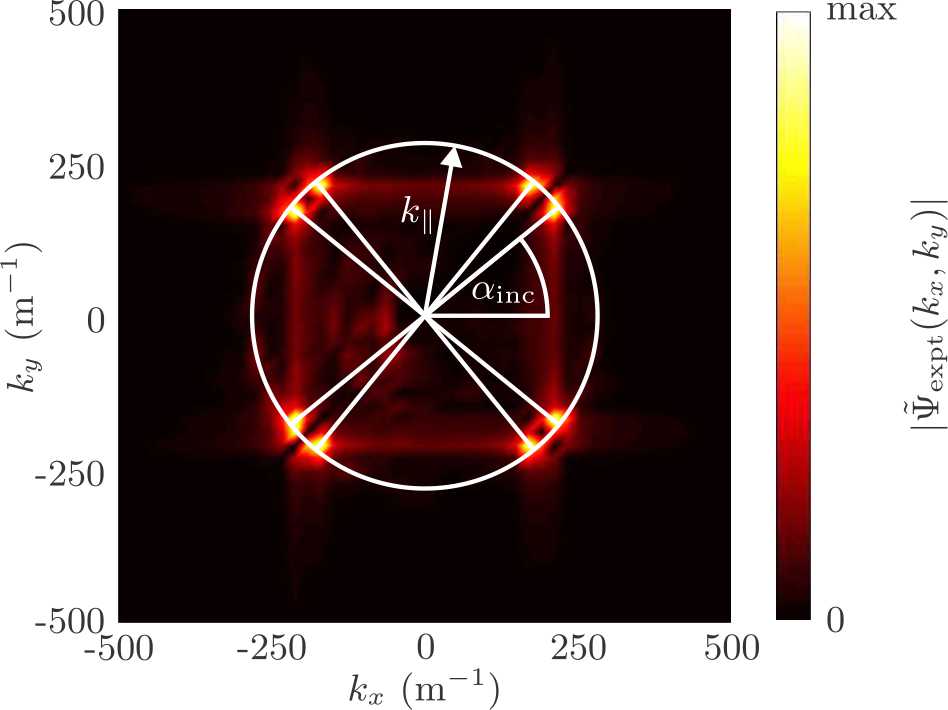}
	\label{sfig:WFgenMS} 
}
\end{center}
\caption{(Color online) Measured wave function \subref{sfig:WFgenCS} and momentum distribution \subref{sfig:WFgenMS} of a TM$_0$ resonance at $6.835$~GHz. The modulus of $\Psiexp(x, y)$ and $\Psitexp(k_x, k_y)$ is shown in false colors, respectively. The straight white lines in panel \subref{sfig:WFgenMS} indicate the eight major momentum components, the white circle indicates their modulus $\kt$, and $\alphainc$ is the angle of incidence of the corresponding family of classical trajectories. Adapted from Ref.~\cite{Bittner2013b}.}
\label{fig:WFgen}
\end{figure}

The WFs were measured on a Cartesian grid covering the whole surface of the resonator with a resolution of $\Delta a = a / 150 \approx 2$~mm [$a / 120 \approx 2.5$~mm in the case of antenna configuration (a)]. An example of a measured WF for a TM$_0$ resonance at $6.835$~GHz is presented in \reffig{sfig:WFgenCS}. The very regular pattern of the WF is due to the relation of the resonant state to a specific set of classical orbits, which can be best understood by considering the corresponding momentum distribution (MD) \cite{Backer1999, Huang2002, Doya2007, Bittner2013b}. It is obtained from the spatial Fourier transform (FT) of the WF inside the resonator,
\begin{equation} \label{eq:defMD} \Psit(k_x, k_y) = \int \limits_{-a/2}^{a/2} dx \int \limits_{-a/2}^{a/2} dy \, \Psi(x, y) e^{-i (k_x x + k_y y)} \, . \end{equation}
The MDs of the measured WFs are calculated using the FFT algorithm and hence have a resolution of $\Delta k_{x, y} = 2 \pi / a$. Note that the MDs can also be directly observed in the far-field of vertical-cavity surface-emitting lasers \cite{Huang2002} or optical fibers \cite{Doya2002, Doya2007}. The MD corresponding to the resonance at $6.835$~GHz is shown in \reffig{sfig:WFgenMS}. It shows a highly symmetric pattern of eight momentum vectors (indicated by the straight white lines) on which it is concentrated. These correspond exactly to one family of orbits defined by a common angle of incidence (indicated as $\alphainc$ in the figure) to which the WF is related. This demonstrates the validity of our ansatz for the model WFs as a superposition of eight plane waves. As can be seen in \reffig{sfig:WFgenMS} the MD also features a structure of faint horizontal and vertical lines connecting the eight major momentum components. They can be considered as artifacts related to the finite size and resolution of the measured WFs for the following two reasons. First, for a finite sampling rate of $x$ the FFT of a complex exponential function $\exp(i k_x x)$ shows a peak of finite width around the momentum $k_x$. Second, the width of the peak depends on how close the momentum $k_x$ is to an integer multiple of the momentum resolution $\Delta k_x$. 

\begin{figure}[tb]
\begin{center}
\subfigure[]{
	\includegraphics[width = 8.4 cm]{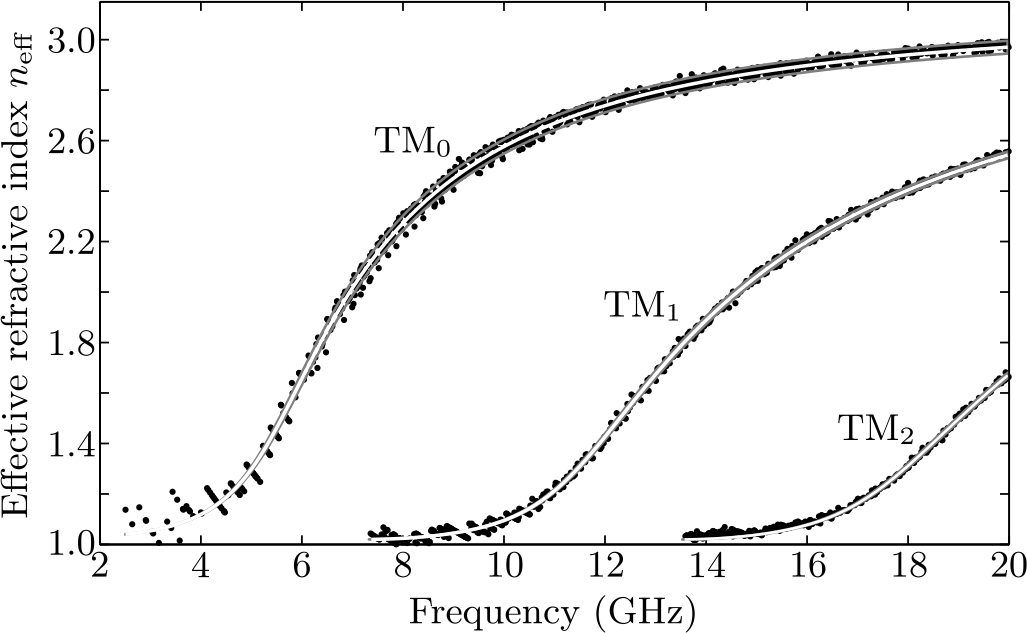}
	\label{sfig:neffFit}
}
\subfigure[]{
	\includegraphics[width = 8.4 cm]{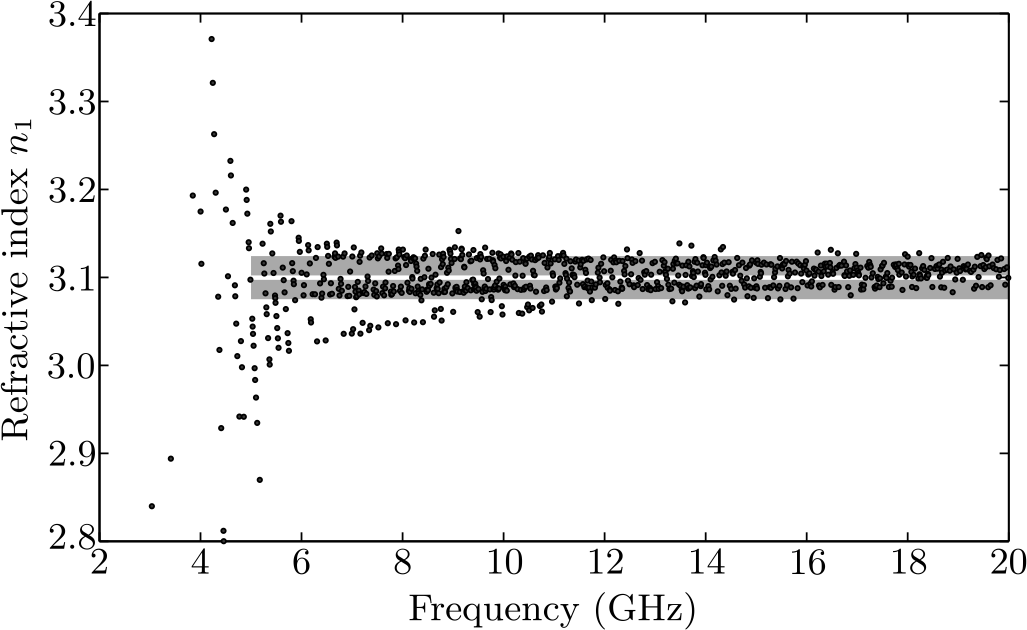}
	\label{sfig:n1Fit}
}
\end{center}
\caption{\subref{sfig:neffFit} Measured effective refractive index. The black data points were obtained from the experimental momentum distributions (see text) and correspond to the three waveguide modes TM$_0$, TM$_1$, and TM$_2$. The white lines were calculated from the theoretical expression for $\neff$, \refeq{eq:neffQuant}, with values for $n_1$ deduced from the experimental data. The gray lines indicate the one sigma error interval of $\neff$. \subref{sfig:n1Fit} Refractive index $n_1$ of the alumina deduced from the measured effective refractive index for the TM$_0$ modes (black points). The white line indicates the mean value of $n_1$ in the range of $5$--$20$~GHz and the gray bar the standard deviation. Note that some data points below $5$~GHz are outside of the displayed range.}
\label{fig:nFit}
\end{figure}

The modulus of the dominant momentum components is $\kt = [k_x^2 + k_y^2]^{1/2} = \neff k$. Therefore the effective refractive index $\neff$ at the resonance frequency can be determined from the distances of the maxima of the MD from the center. The thus-determined effective refractive index is related to the TM$_0$ slab waveguide modes and corresponds to the first branch of data points displayed in \reffig{sfig:neffFit}. The data points scatter only slightly around the white line, which is the theoretical curve of $\neff$ given by \refeq{eq:neffQuant} for $n_1 = 3.1$. This value for the refractive index of the alumina was obtained as follows: The measured values of $\neff$ were inserted into \refeq{eq:neffQuant}, which was subsequently solved for $n_1$. These values of $n_1$ are shown in \reffig{sfig:n1Fit}. In the range of $5$--$20$~GHz, the data points for $n_1$ scatter only little around the mean value of
\begin{equation} \label{eq:fittedN1} \left< n_1 \right> = 3.100 \pm 0.025 \, . \end{equation}
The data show no significant frequency dependence, and therefore we can neglect any dispersion of the refractive index. The outliers below $5$~GHz are due to the finite resolution of the MDs that has a particularly strong influence for small values of $\kt$, respectively, $\neff$. The value for $n_1$ given in \refeq{eq:fittedN1} was then used to calculate the white line shown in \reffig{sfig:neffFit} via \refeq{eq:neffQuant}. These values for $n_1$ and $\neff$, respectively, are used in all further calculations throughout this article. The error band of the calculated $\neff$ corresponding to the standard deviation of $n_1$ is indicated by the gray lines. The sole unknown parameter here is $n_1$ since the thicknesses of the alumina plate and the foam, $b$ and $\afoam$, are known with high precision. The refractive index of the foam, $n_2$, is also not known precisely, but the dependence of $n_1$ on $n_2$ is negligible, so it could be considered as a constant. It should be noted that $\neff$ can be determined from the measured WFs even in a regime of strongly overlapping resonances (i.e., also above $10$~GHz) since the phase velocity in the resonator slab is identical for all resonances of the same polarization and $z$ excitation.

\begin{figure*}[tb]
\begin{center}
\includegraphics[width = 12 cm]{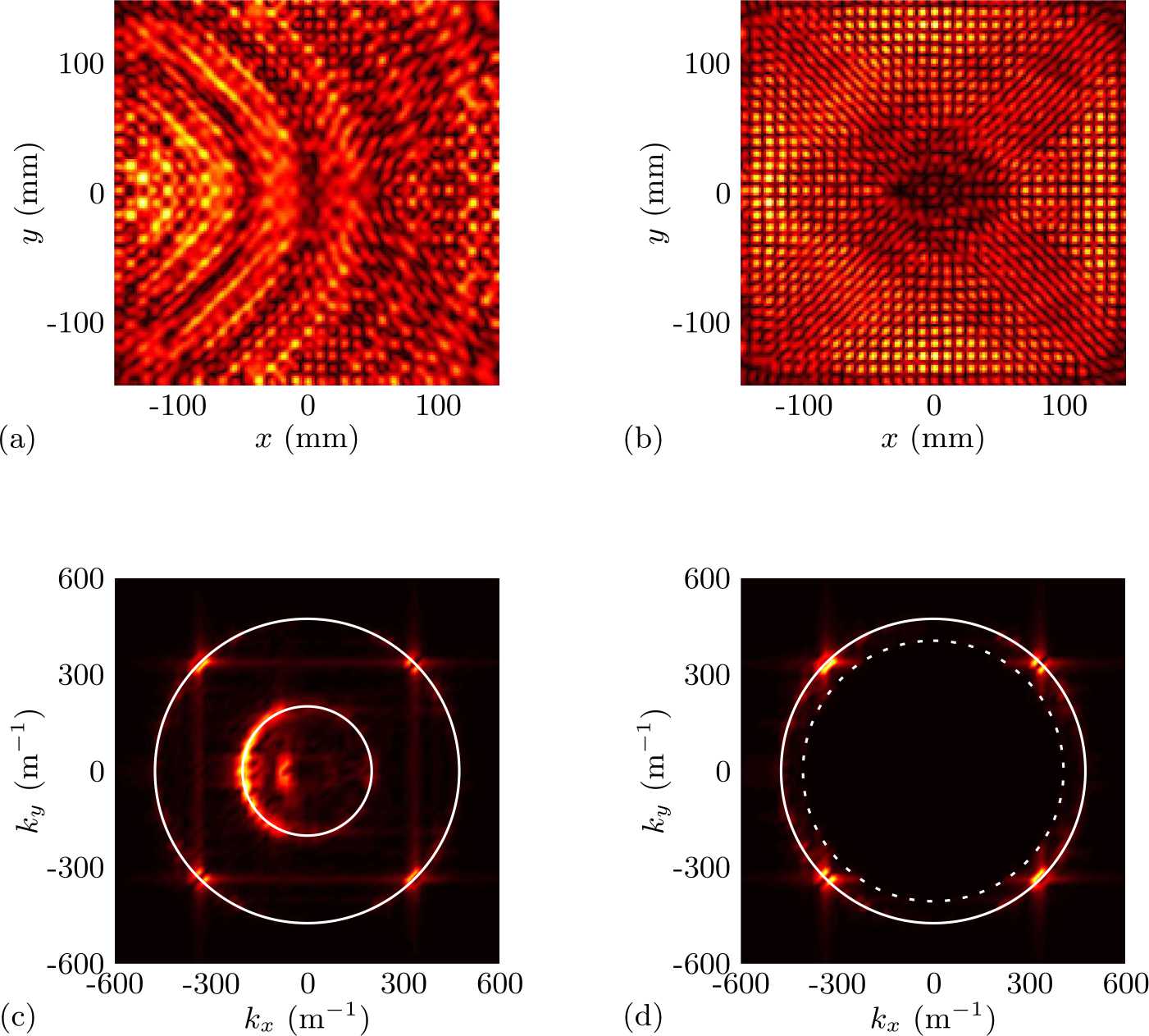}
\end{center}
\caption{(Color online) (a) Measured WF of a TM$_0$ resonance at $9.146$~GHz. The excitation antenna was placed at the midpoint of the left side. See \reffig{sfig:WFgenCS} for the color scale. (b) Corresponding filtered WF. (c) Corresponding momentum distribution. The outer white circle indicates $k \neff(\mathrm{TM}_0)$, and the inner one $k \neff(\mathrm{TM}_1)$. See \reffig{sfig:WFgenMS} for the color scale. (d) Filtered momentum distribution. The dashed white circle indicates the filter radius $\kfilt$.}
\label{fig:zetaFilt}
\end{figure*}

The determination of the MDs turned out to be also useful for the assignment to a symmetry class and finally a model WF of measured WFs that do not exhibit a clear structure. An example for such a WF, that of a TM$_0$ resonance at $9.146$~GHz, is presented in \reffig{fig:zetaFilt}(a). Its pattern is overlain by a structure of concentric circles centered around the position of the excitation antenna at the midpoint of the left edge of the resonator. The cause of this pattern becomes apparent in the corresponding MD shown in \reffig{fig:zetaFilt}(c). It features eight points of high intensity lying on a circle with radius $k \neff(\mathrm{TM}_0)$, indicated by the outer white circle, like the MD shown in \reffig{sfig:WFgenMS}. In contrast to the latter, however, the MD in \reffig{fig:zetaFilt}(c) also features other significant contributions to the MD. These are concentrated on the inner white circle in \reffig{fig:zetaFilt}(c). They stem from propagating, nonresonant TM$_1$ slab waveguide modes, and the radius of this circle is $\kt = k \neff(\mathrm{TM}_1)$. Further contributions inside the inner circle are attributed to direct transmission between the antennas. The measured WF in \reffig{fig:zetaFilt}(a) is hence a superposition of the resonant TM$_0$ mode and propagating TM$_1$ waves. Such an interference effect was not observed in \reffig{sfig:WFgenCS} since the corresponding mode is below the cut-off frequency of the TM$_1$ mode, $\fcrit(\mathrm{TM}_1) \approx 7.34$~GHz. In general, however, the wire antennas excite waves in all available TM waveguide modes, where the cut-off frequency of the TM$_2$ modes is $\fcrit(\mathrm{TM}_2) \approx 13.53$~GHz. Accordingly, also the effective refractive indices of the TM$_{1, 2}$ modes can be measured and correspond to the second and third branch of data points, respectively, in \reffig{sfig:neffFit}. The resulting values of $n_1$ for the TM$_1$ mode in the range of $11$--$20$~GHz and for the TM$_2$ mode in the range of $17$--$20$~GHz are $\left< n_1 \right> = 3.098 \pm 0.011$ and $\left< n_1 \right> = 3.093 \pm 0.007$, respectively. These values are in very good agreement with the value deduced from the TM$_0$ modes given in \refeq{eq:fittedN1}.

The excitation of and hence the interference between different waveguide modes in the resonator is unfortunately unavoidable with a simple antenna design as was used for the experiment presented in this article. Furthermore, the coupling of a wire antenna situated above or below the resonator to the higher TM modes is generally stronger than to the TM$_0$ mode. The reason is that the decay lengths of the higher-excited modes are larger since their effective refractive indices are smaller than that of the TM$_0$ mode [see \refeq{eq:decayLength} and \reffig{fig:nFit}(a)]. Therefore, the field distributions of the TM$_0$ modes that we are interested in are increasingly obscured at higher frequencies. Since, however, the different waveguide modes are well separated in momentum space the higher TM modes can be filtered out of the measured WFs with relative ease. The part of the MD inside a circle with radius $\kfilt = k \left[ 0.75 \, \neff(\mathrm{TM}_0) + 0.25 \, \neff(\mathrm{TM}_1) \right]$ is simply set to zero. In \reffig{fig:zetaFilt}(d), the boundary of this circle is indicated by the dashed white circle. The filter radius is chosen relatively close to $\kt$ to ensure that all contributions of other modes are cut out without affecting the field distribution originating from the resonant mode itself. The filtered WF shown in \reffig{fig:zetaFilt}(b) was obtained by computing the inverse FT. The concentric circles around the excitation antenna have disappeared and predominantly the field distribution of the resonant mode itself remains.

This filtering technique enables us to procure high-quality data in frequency regimes where this would normally be impossible. It should be noted that only filtered WFs are shown and used in the following analyses. Also the results presented in Ref.~\cite{Bittner2013b} were based exclusively on filtered WFs. Evidently, the same filter technique can be used to isolate the TM$_{\zeta > 0}$ contributions as well. Indeed, also some TM$_1$ resonant states were found and could be assigned to model modes (not shown here).

In conclusion, the analysis of the MDs of the resonant states leads to a profound understanding of the measured WFs. It also demonstrates directly the existence of different waveguide modes in a thin resonator and the validity of the calculation of the corresponding effective refractive indices. It should be emphasized that the techniques described in this section can be applied to any flat microwave resonator. In particular, the determination of $\neff$ and successively of $n_1$ directly from the measured WFs was used, e.g., to validate the values of the refractive indices used in Refs.~\cite{Bittner2009, Bittner2012a, Bittner2013b}.

\section{Comparison of experimental data and model calculations} \label{sec:expVsMod}

\subsection{Identification with model modes} \label{ssec:modID}

\begin{figure*}[tb]
\begin{center}
\includegraphics[width = 15.0 cm]{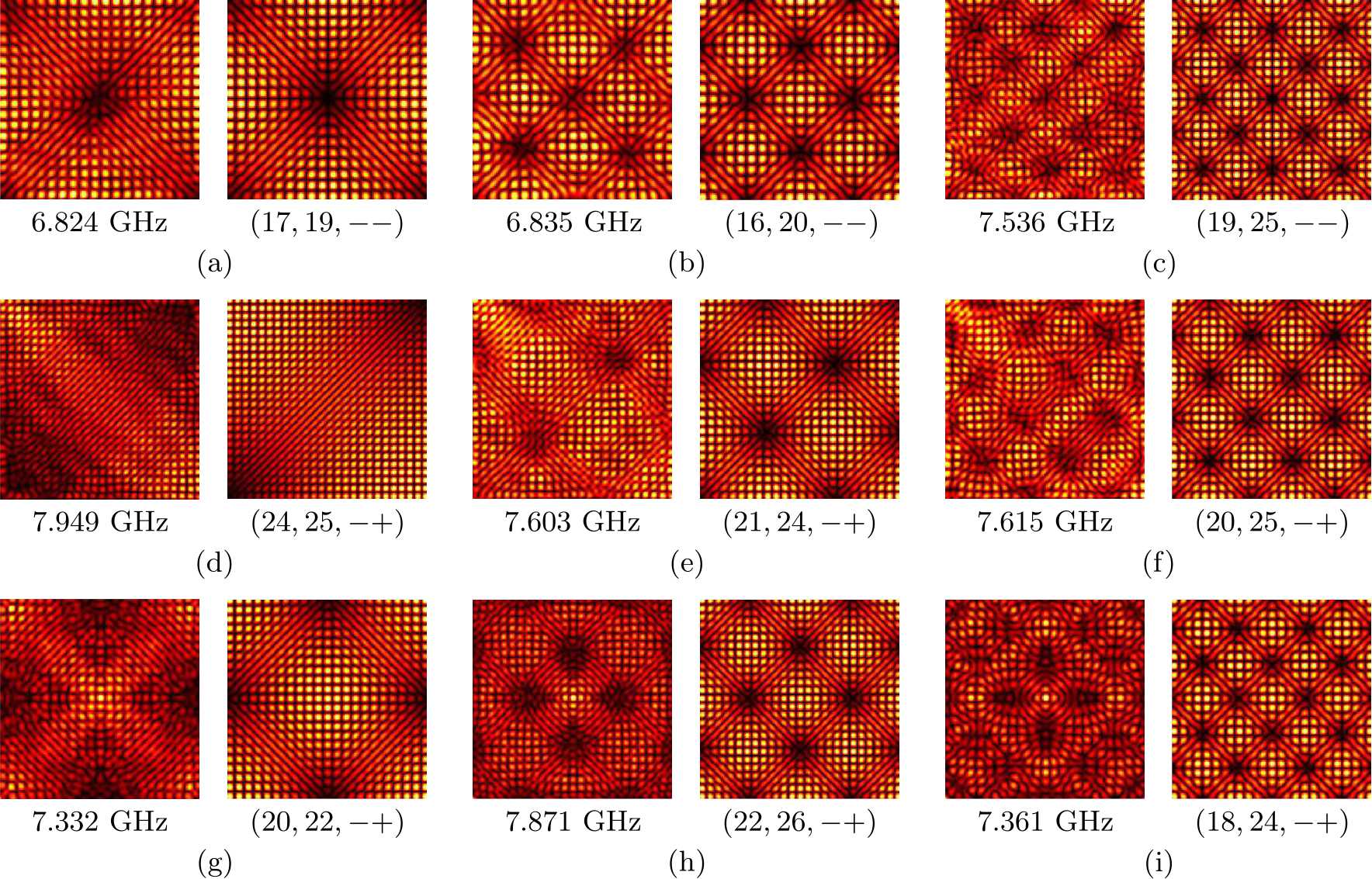}
\end{center}
\caption{(Color online) Measured WFs (left subpanels) and corresponding model WFs (right subpanels). See \reffig{sfig:WFgenCS} for the color scale. The corresponding measured resonance frequencies (left subpanels) and quantum numbers (right subpanels) are indicated. The WFs presented in (a)--(c) have $(--)$ symmetry and were excited by the fixed antenna placed $a/4$ above the lower left corner, the WFs in (d)--(f) have $(-+)$ symmetry and were excited by the fixed antenna placed at the upper left corner, and the WFs in (g)--(i) have $(++)$ symmetry and were excited by the fixed antenna placed in the middle of the resonator.}
\label{fig:WFexmpls}
\end{figure*}

Figure \ref{fig:WFexmpls} shows several examples of measured WFs (left subpanels) and the corresponding model WFs (right subpanels). The latter could be un\-am\-bi\-guous\-ly identified by calculating the overlaps $|\left< \Psiexp(f) \middle| \Psimod(m_x, m_y, s_1 s_2) \right>|^2$ with several trial WFs. Generally, this was possible if the overlap with just one model WF was greater than $40\%$ while the overlaps with all other eligible model functions were negligible (cf.\ Ref.~\cite{Bittner2013b}). The overlaps in the cases presented in \reffig{fig:WFexmpls} are in the range of $60\%$ to $85\%$. On average, the overlaps are somewhat larger for the $(--)$ modes than for the $(-+)$ and the $(++)$ modes, which explains why the measured and model modes shown in the first row of Fig. 6 exhibit a better visual agreement than those shown in the other two rows. Some of the TM$_0$ modes indicated by the arrows in \reffig{fig:freqSpectra} could not be clearly assigned to one model WF because of accidental degeneracies (or near degeneracies) with other modes. This effectively limited the frequency range in which the measured modes could be unambiguously related to model WFs to $\fmax = 10$~GHz. In the following we restrict our discussion to those that were clearly identified as explained above, consisting of $166$ resonant states in total.

It is instructive to define a different set of quantum numbers $(m, p)$ via $m = m_x + m_y$ and $p = |m_y - m_x|/2$. We call $m$ the longitudinal and $p$ the transverse quantum number because they correspond to the momentum components parallel and perpendicular to the periodic orbit channel of the diamond PO \cite{Bittner2011a}. The possible values of the quantum numbers $m$ and $p$ depend on the symmetry class of a mode (see also \reftab{tab:dlmWFs}). For $(--)$ modes, $m$ is even and $p = 1, 2, 3, \dots$. The WFs presented in Figs.~\ref{fig:WFexmpls}(a)--\ref{fig:WFexmpls}(c) have transverse quantum numbers $p=1$, $2$, and $3$, respectively. For $(-+)$ and $(+-)$ modes, $m$ is odd and $p = 0.5, 1.5, 2.5, \dots$. The WFs in Figs.~\ref{fig:WFexmpls}(d)--\ref{fig:WFexmpls}(f) have transverse quantum numbers $p = 0.5$, $1.5$, and $2.5$, respectively. For $(++)$ modes, finally, $m$ is even and the transverse quantum number can take the values of $p = 0, 1, 2, \dots$. The WFs in Figs.~\ref{fig:WFexmpls}(g)--\ref{fig:WFexmpls}(i) have $p=1$, $2$, and $3$, respectively. The modes with $p = 0$, i.e., $m_x = m_y$, are a special case (not shown here). They can be described only by a superposition of two model WFs with $(++)$ symmetry. The reason is a non-negligible coupling between the $(m_x, m_y, ++)$ and the $(m_x-2, m_y+2, ++)$ modes \cite{Bittner2013b, Wiersig2006}.

Some simplified models \cite{Guo2003, Lebental2007, Bittner2011a} concentrate on modes related to families of orbits close to the diamond PO since those modes are usually the most prominent ones in microlaser and -cavity experiments. In these models a wave is embedded in the periodic orbit channel parallel to the family of the diamond PO that is then folded back to obtain the actual model WF \cite{Bogomolny2006, Bittner2011a}. The embedded wave is characterized by one momentum vector component parallel and one perpendicular to the PO, $k_\chi$ and $k_\eta$, respectively. These are quantized via \cite{Lebental2007, Bittner2011a}
\begin{equation} \lpo k_\chi = 2 \pi m + 8 \delta \end{equation}
and 
\begin{equation} \frac{\lpo}{4} k_\eta = p \pi, \end{equation}
where $\lpo = 2 \sqrt{2} a$ is the length of the diamond PO. The term $8 \delta$ corresponds to the phase shifts at the four reflections with $\alphainc = 45^\circ$, where
\begin{equation} \label{eq:quantCondLong} \delta = - \frac{1}{2} \arg[r(45^\circ)] = \arctan \left( \sqrt{\neff^2-2} / \neff \right) \, . \end{equation}
This model gives the resonance frequency as $\ftheo = c [k_\chi^2 + k_\eta^2]^{1/2} / (2 \pi \neff)$. Actually, for $\alphainc = 45^\circ$ it yields the same resonance frequencies and WFs as the model presented in \refsec{ssec:rayModel}. For small $p$, the angle of incidence $\alphainc$ does not deviate strongly from $45^\circ$ and hence in these cases both models predict almost the same resonance frequencies and the model WFs are practically indistinguishable. This is the case, for example, for the modes presented in Figs.~\ref{fig:WFexmpls}(a)--\ref{fig:WFexmpls}(c) that are associated to trajectories with angles of incidence $\alphainc = 42.0^\circ$, $39.1^\circ$, and $37.6^\circ$, respectively. The deviations between the two models increase with $p$, i.e., with that of $\alphainc$ from $45^\circ$, until the diamond-PO-based model is no longer applicable. In summary, the diamond-PO-based models are contained as a limiting case in our ray-based model which, in contrast to the former, is valid for all types of modes.

\subsection{Symmetry properties of the measured wave functions} \label{ssec:symProp}

\begin{table}[tb]
\caption{Overview of the set of unambiguously identified modes. The columns give the number of modes found with respect to their symmetry class and the rows the number with respect to the position of the excitation antenna. See the insets of \reffig{fig:freqSpectra} for an illustration of the antenna positions.}
\label{tab:WFsymVSant}
\vspace{3 mm}
\begin{center}
\begin{tabular}{l|c||c|c|c}
\hline
\hline
Antenna position & Total & $(--)$ & $(-+)$ & $(++)$ \\
\hline
a/4 from corner & 27 & 25 & - & 2 \\
Corner & 62 & - & 56 & 6 \\
Middle & 53 & - & - & 53 \\
Midpoint of side & 24 & 23 & - & 1 \\
\hline
Total & 166 & 48 & 56 & 62 \\
\hline
\hline
\end{tabular}
\end{center}
\end{table}

The symmetry of the resonant modes that are excited strongly depend on the position of the excitation antenna as discussed in \refsec{sec:freqSpec}. Table \ref{tab:WFsymVSant} gives an overview of the number of modes with a given symmetry that were unambiguously identified for the different positions of the excitation antenna. The set contains no modes with $(+-)$ symmetry since these are identical to the modes with $(-+)$ symmetry except for a rotation by $90^\circ$ and hence measurements with a corresponding antenna position were omitted. While the WFs are perfectly (anti-)symmetric with respect to the various symmetry axes in theory, the measured WFs do not exhibit a perfect symmetry due to unavoidable experimental imperfections. These can, in general, be perturbations of the resonator geometry or, in our case, inaccuracies in the positioning of the excitation antenna and small contributions from nearby resonances with different symmetries. 

The actual degree of symmetry of a measured WF can be quantified by the symmetry ratios $|C_{s_1 s_2}|^2$ defined via
\begin{equation} C_{s_1 s_2} = \left< \Psiexp^{(s_1 s_2)} \middle| \Psiexp \right> \end{equation}
where $\Psiexp^{(s_1 s_2)}$ is the part of $\Psiexp$ with $(s_1 s_2)$ symmetry, given by
\begin{equation} \Psi^{(s_1 s_2)} = \frac{1}{4} (\mathbbm{1} + s_1 \Pone) (\mathbbm{1} + s_2 \Ptwo) \Psi \, . \end{equation}
The operator $\Pone$ ($\Ptwo$) mirrors a WF with respect to the $x=y$ ($x=-y$) axis and $\mathbbm{1}$ is the identity operator. By definition, $|C_{--}|^2 + |C_{-+}|^2 + |C_{+-}|^2 + |C_{++}|^2 = 1$.

\begin{figure}[tb]
\begin{center}
\subfigure[]{
	\includegraphics[width = 8.0 cm]{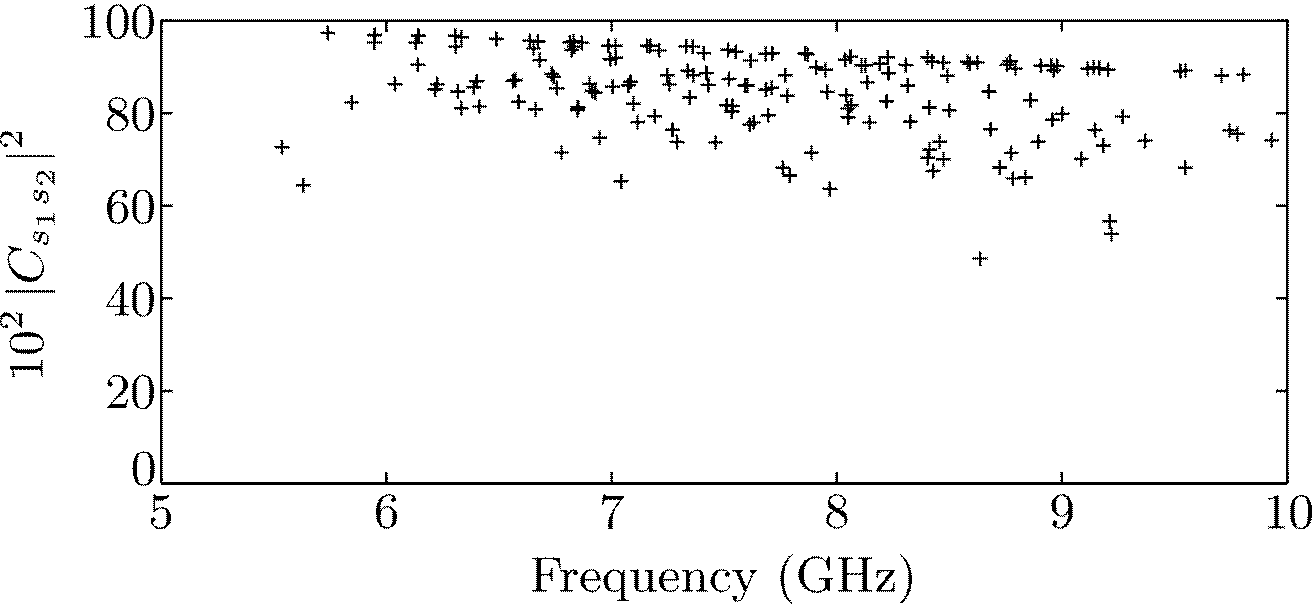}
	\label{sfig:symRaMax}
}
\subfigure[]{
	\includegraphics[width = 8.0 cm]{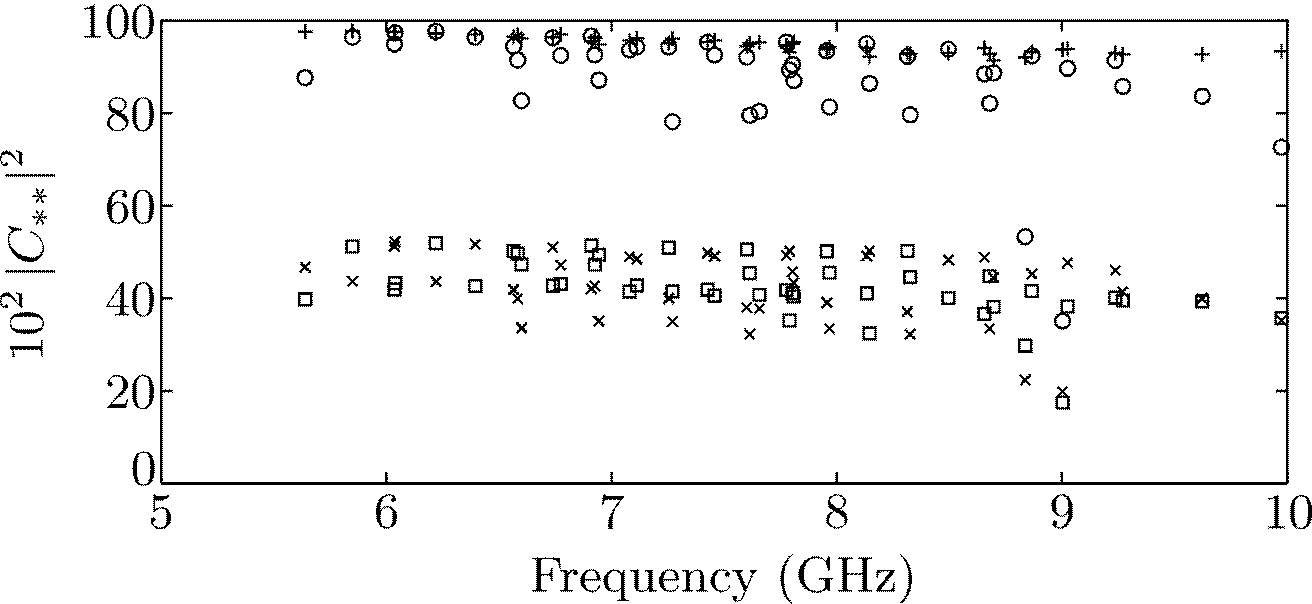}
	\label{sfig:symRaVH}
}
\end{center}
\caption{\subref{sfig:symRaMax} Symmetry ratios $|C_{s_1 s_2}|^2$ of resonant states unambiguously assigned to $(m_x, m_y, s_1 s_2)$ modes versus the frequency. The data set comprises modes belonging to the $A_{1,2}$ and $B_{1,2}$ representations as well as to the $E$ representation with $(-+)$ symmetry, the latter being excited by an antenna at the corner of the resonator. \subref{sfig:symRaVH} Symmetry ratios of the TM$_0$ modes belonging to the $E$ representation where the excitation antenna was placed at the midpoint of the left edge [cf.\ \reffig{fig:freqSpectra}(d)]. The different symmetry ratios are marked by $+$ for $|C_{y+}|^2$, $\circ$ for $|C_{x-}|^2$, $\times$ for $|C_{-+}|^2$, and $\square$ for $|C_{+-}|^2$.}
\label{fig:symRatios}
\end{figure}

The values of $|C_{s_1 s_2}|^2$ of the considered $166$ modes are shown in \reffig{sfig:symRaMax}. The values are typically in the range of $75\%$ to $95\%$. The maximal values obtained decrease with increasing frequency because the resonant states become more sensitive to geometric deviations with decreasing wavelength. The overlap of a measured WF with a model WF having $(s_1 s_2)$ symmetry must of course be smaller or equal to the corresponding symmetry ratio $|C_{s_1 s_2}|^2$. This additionally impedes the clear identification of modes with increasing frequency. 

The symmetry ratios $|C_{x s_x}|^2$ and $|C_{y s_y}|^2$ with respect to the vertical and horizontal axes, respectively, can be calculated in the same manner, where $s_x, s_y \in \{ +, - \}$ and the corresponding (anti-)symmetric parts of the WFs are
\begin{equation} \Psi^{(x s_x)} = \frac{1}{2} [ \Psi(x, y) + s_x \Psi(-x, y) ] \end{equation}
and 
\begin{equation} \Psi^{(y s_y)} = \frac{1}{2} [ \Psi(x, y) + s_y \Psi(x, -y) ] \, . \end{equation}
Similarly, $|C_{x+}|^2 + |C_{x-}|^2 = 1$ and $|C_{y+}|^2 + |C_{y-}|^2 = 1$. The symmetry ratios $|C_{x+}|^2$ and $|C_{y+}|^2$ of the modes assigned to the $A_1$ and $B_2$ representations as well as the symmetry ratios $|C_{x-}|^2$ and $|C_{y-}|^2$ for those belonging to the $A_2$ and $B_1$ representations were in the range of $90$--$100\%$ when placing the excitation antenna on the horizontal or vertical symmetry axis (i.e., in the middle of the square or at the midpoint of an edge) and a bit smaller ($75$--$90\%$) otherwise. So the measured WFs of the modes with $(--)$ and $(++)$ symmetry exhibit the expected symmetries to a high degree, regardless of the position of the excitation antenna. 

The case of the modes belonging to the $E$ representation is more complicated. When the excitation antenna was placed at a corner of the resonator [see inset of  \reffig{fig:freqSpectra}(b)], they exhibited a high degree of $(-+)$ symmetry as shown in \reffig{sfig:symRaMax}. In contrast, the values of $|C_{x\pm}|^2$ and $|C_{y\pm}|^2$ were around $50\%$ (not shown), i.e., they did not have a well-defined symmetry with respect to the vertical and horizontal axes as indicated in the last two rows of \reftab{tab:dlmWFs} and, consequently, could be identified with these model WFs. When the excitation antenna was put at the midpoint of the left edge [see inset in \reffig{fig:freqSpectra}(d)], the situation was reversed. This is exemplified in \reffig{sfig:symRaVH} where the different symmetry ratios of the modes belonging to the $E$ representation\footnote{These modes are not part of the data set of the $166$ unambiguously identified modes.} are shown. They all have a high symmetry ratio $|C_{y+}|^2$ in the range of $90$--$100\%$. The symmetry ratios $|C_{x-}|^2$ have a similarly high level as is expected since modes of the $E$ representation that are symmetric (antisymmetric) with respect to one symmetry axis must be antisymmetric (symmetric) with respect to the perpendicular one. In contrast, the symmetry ratios $|C_{-+}|^2$ ($\times$) and $|C_{+-}|^2$ are around $50\%$. Consequently, these modes can neither be identified with the $(+-)$ nor with the $(-+)$ model WFs listed for the $E$ representation in \reftab{tab:dlmWFs}. In accordance with this observation, the overlaps of the measured WFs with the model WF having $(-+)$, respectively, $(+-)$ symmetry are approximately equal. Similarly, when the excitation antenna was placed $a/4$ away from a corner, i.e., not on any of the symmetry axes, the measured WFs belonging to the $E$ representation did not exhibit any well-defined symmetry since none was induced by the position of the antenna, that is, they also could not be identified with one of the model WFs given in the last two rows of \reftab{tab:dlmWFs}. Therefore in order to observe modes exhibiting clear $(-+)$ [or $(+-)$] symmetry, one antenna had to be positioned at a corner of the resonator (cf.\ \reftab{tab:WFsymVSant}). In conclusion, the symmetry properties of the modes belonging to the $E$ representation are determined solely by the position of the excitation antenna. The reason is that these modes come in degenerate pairs and there is hence a degree of freedom as concerns the symmetry of their WFs. In contrast, the nondegenerate $(--)$ and $(++)$ modes exhibit their symmetries independently of the antenna position, i.e., they are only due to the geometry of the resonator itself.

\subsection{Review of the experimental data} \label{ssec:dataReview}

\begin{figure*}[tb]
\begin{center}
\includegraphics[width = 12.0 cm]{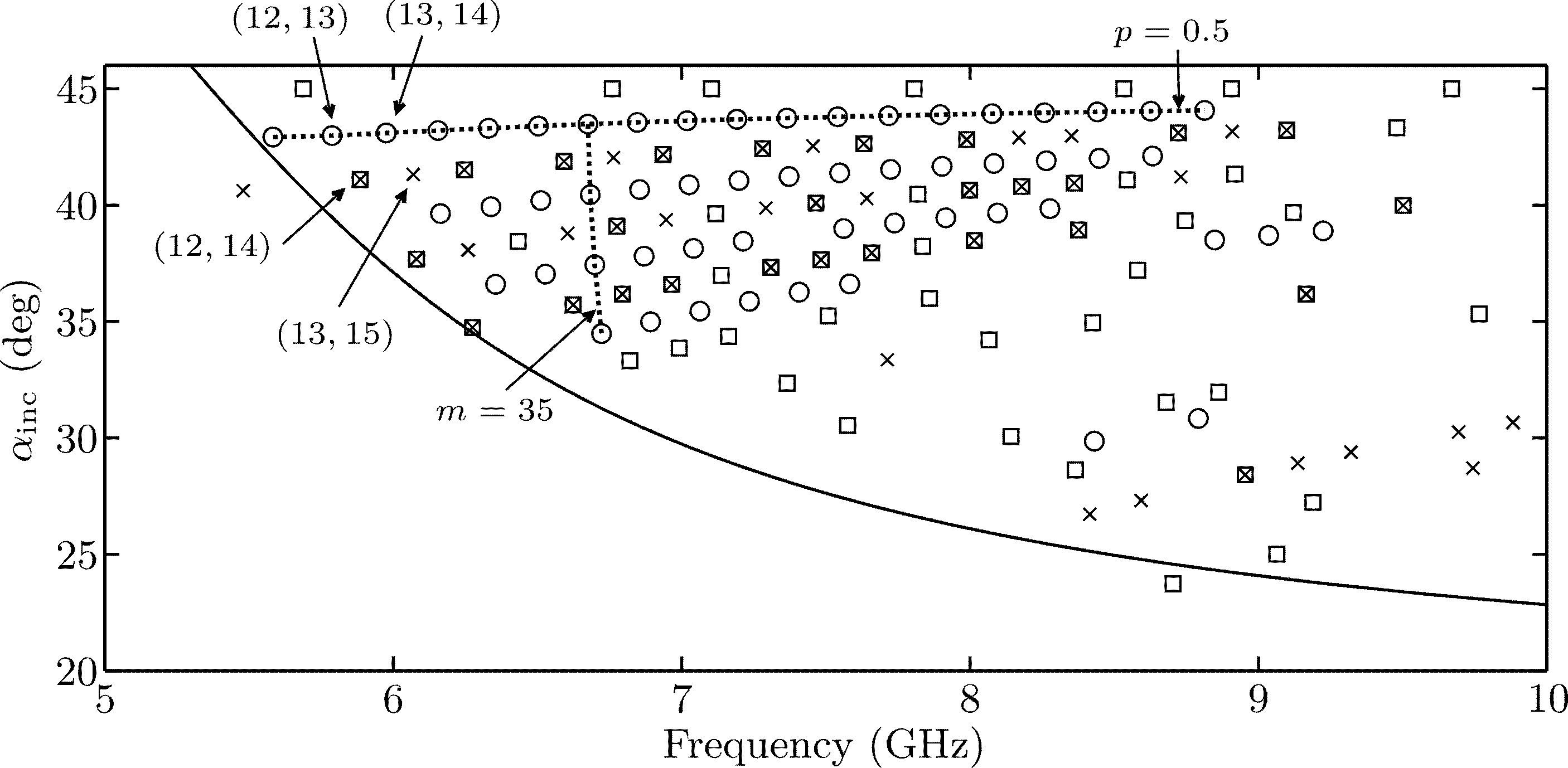}
\end{center}
\caption{Overview of the identified measured resonances. The angle of incidence $\alphainc$ of the corresponding set of classical trajectories is plotted versus the measured resonance frequency. The different symbols indicate the symmetry class, marked by $\square$ for $(++)$, $\times$ for $(--)$, and $\circ$ for $(-+)$. The quantum numbers $(m_x, m_y)$ of some resonances are indicated. The horizontal dotted line indicates the series of resonances with transverse quantum number $p = 0.5$, and the vertical dotted line indicates the series with longitudinal quantum number $m = 35$. The solid line indicates the critical angle for TIR, $\alphacrit$.}
\label{fig:alphaVsFreq}
\end{figure*}

The angle of incidence is a constant of motion that defines the classical tori. An overview of the set of $166$ resonances listed in \reftab{tab:WFsymVSant} is presented in \reffig{fig:alphaVsFreq}. Each of the associated modes corresponds to a certain set of classical trajectories. The corresponding angles of incidence $\alphainc$ are given as a function of the resonance frequency. The different symbols correspond to the different symmetry classes. It should be noted that the $(++)$ and the $(--)$ modes are not degenerate although their resonance frequencies seem to be identical on the scale of the figure. The modes form a regular, gridlike pattern in this diagram. This can be regarded as an indication that the dielectric square resonator behaves like an integrable system \cite{Peres1984}. The gridlike structure is exemplified by the horizontal dotted line that indicates a series of modes with fixed transverse quantum number $p = 0.5$ and by the vertical dotted line that indicates a series of modes with fixed longitudinal quantum number $m = 35$. The series with $\alphainc = 45^\circ$ forming the top line of modes consists of those with $m_x = m_y$, i.e., $p=0$. They all showed coupling to the neighboring $(m_x-2, m_x+2, ++)$ modes as described in Ref.~\cite{Bittner2013b}. It should be noted that a series of modes with constant transverse quantum number $p$ close to $0$ has a free spectral range (FSR) of $\Delta k = k_{m+2, p} - k_{m, p} \approx \sqrt{2} \pi / (a \neff)$. In experiments with optical microcavities or -lasers, often a series of resonances with half this FSR is observed \cite{Poon2001, Chern2004, Lee2004a, Lebental2007}. Such a series must therefore consist alternately of modes with either $(--)$ or $(++)$ symmetry and of modes with $(-+)$ or $(+-)$ symmetry, i.e., of modes belonging to two families with different $p$ values. 

There are many vacancies in the diagram since not all modes could be found experimentally, especially above $9$~GHz, due to the deterioration of the data quality. Furthermore, all observed modes (with two exceptions) have an angle of incidence that is larger than the critical angle $\alphacrit$ indicated by the solid line in \reffig{fig:alphaVsFreq}. The model of course also predicts modes that are not confined by TIR; however, these cannot be observed in an experiment with a passive resonator since refractive losses render them too short lived. This is also the reason why modes could only be clearly identified for frequencies above $\fmin = 5.5$~GHz, which is approximately the frequency at which $\neff$ reaches the value of $\sqrt{2}$ and hence $\alphacrit$ drops below $45^\circ$ [cf.\ \reffig{sfig:neffFit}]. Furthermore, while for frequencies above and close to $\fmin$ only one or two series of modes with constant $p$ lie above the critical angle and are observed, more series with higher transverse quantum number appear with increasing frequency since $\neff$ grows and, consequently, $\alphacrit$ decreases. This effect leads to the increase in the resonance density observed in the measured spectra in \reffig{fig:freqSpectra}. 

It should be noted that the model predicts an infinite lifetime for all modes with $\alphainc \geq \alphacrit$, i.e., in particular for those that are observed experimentally. The reason for this is that 
the underlying calculations are based on the Fresnel coefficients for an infinite dielectric interface that yield total reflection for $\alphainc \geq \alphacrit$. In reality, however, all modes have a finite lifetime due to radiative losses. In order to account for this, the reflection coefficients must be modified in a nontrivial manner for an interface with finite length $a$ as in the case of the square, leading to finite losses also above $\alphacrit$. This will be the subject of a future publication \cite{BittnerPrep}. The experimentally measured resonance widths, on the other hand, stem not only from radiative losses but also from other mechanisms such as absorption in the alumina and coupling out by the antennas. These latter loss mechanisms are, unfortunately, dominant, and hence no reliable information on the radiative losses of the modes confined by TIR can be extracted from the experimental data. 

\begin{figure}[tb]
\begin{center}
\includegraphics[width = 8.4 cm]{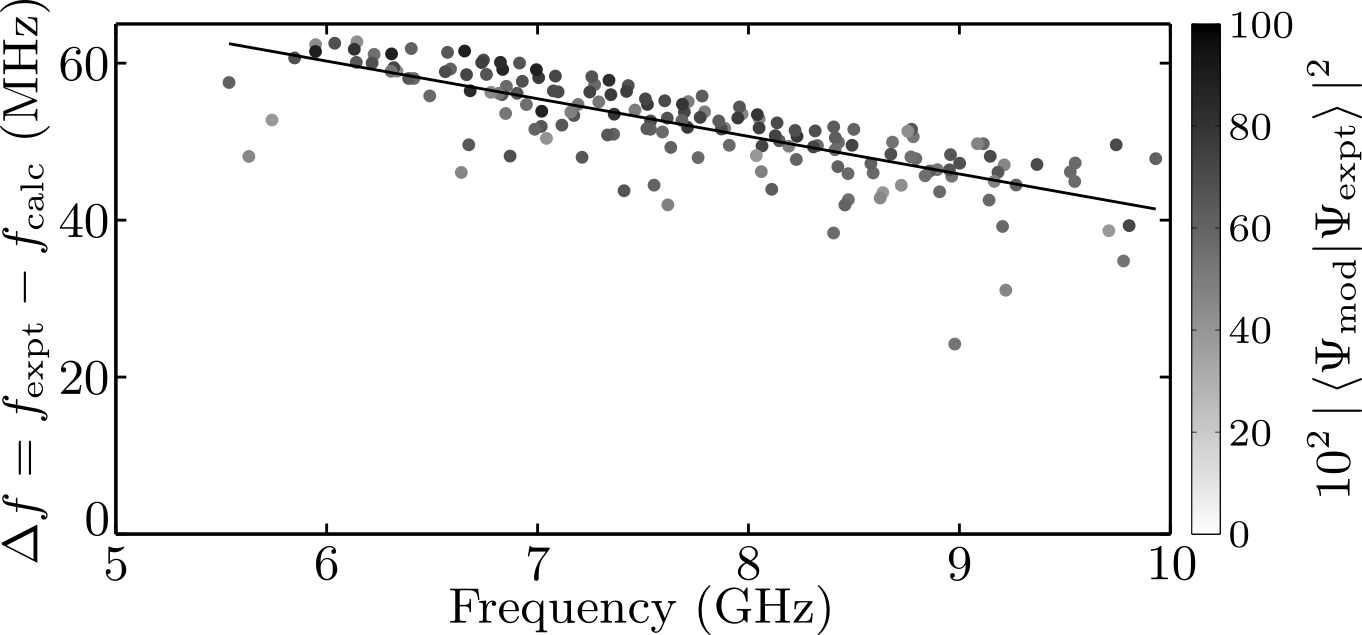}
\end{center}
\caption{Difference between measured and calculated resonance frequencies versus the measured frequency. The gray scale indicates the overlap between the experimental and model WFs. The solid line is a linear fit to $\Delta f$.}
\label{fig:freqComp}
\end{figure}

The difference between the measured resonance frequencies $\fexp$ and those calculated according to the model, $\ftheo$, is shown in \reffig{fig:freqComp}. The relative deviations are in the range of $0.4$--$1.0\%$ and decrease with increasing frequency. There are two possible reasons for these relatively small but nonetheless significant deviations. First, the ray-based model does not provide an exact but only an approximate solution of the Helmholtz equation. It can be expected, though, that it is more precise in the short-wavelength limit due to its semiclassical nature as evidenced by the data in \reffig{fig:freqComp}. Second, the system studied here is approximated as a two-dimensional one by means of the effective refractive index model (see \refsec{ssec:neffApp}). It is known that this approximation can predict resonance frequencies only with limited precision \cite{Bittner2009}, even though it was shown in \refsec{sec:fieldDistr} that the propagation of waves inside the resonator is described with high precision by the $\neff$ model. Indeed, the observed deviations are of the same order of magnitude as those found in Ref.~\cite{Bittner2009} and show the same qualitative behavior. It is surmised that the precision of the effective refractive index approximation could be improved by taking into account the finite height of the cavity side walls. Note that this must be distinguished from the modification of the reflection coefficients due to the finite extension of the dielectric interface in the plane of the resonator mentioned in the previous paragraph. This would correspond to a modification of the reflection coefficients used in \refeq{eq:quantCond}, and the corresponding change of the reflection phases could account for the frequency deviations. In the case studied here, however, it is not clear to what extent these two approximations contribute to the deviations between measured and calculated resonance frequencies each. The difference decreases approximately linearly with increasing frequency. The solid line is a linear fit to $\Delta f = \fexp - \ftheo$,
\begin{equation} \label{eq:freqDevFit} \Delta \ffit = A - B \fexp \, , \end{equation}
where each data point was weighted by the overlap between the model and the measured WF. The fit parameters are $A = (89.0 \pm 0.6)$~MHz and $B = (4.79 \pm 0.01)$~MHz/GHz. It should be noted that a linear fit was chosen for the sake of convenience and because it describes the data well. In reality, we expect that the frequency deviations tend to zero in an asymptotic manner for $f \rightarrow \infty$.

\section{Length spectrum and trace formula} \label{sec:lspect}
While the resonant states of the dielectric square resonator are associated with classical tori that generally consist of nonperiodic trajectories, the spectrum of the resonator can nonetheless be associated with the POs of the classical square billiard. This connection is expressed by a trace formula. It connects the density of states (DOS) of wave-dynamical systems with the POs of the corresponding classical (or ray-dynamical) system \cite{Gutzwiller1970, Gutzwiller1971, Brack2003}. Recently, it has also been applied to open dielectric resonators \cite{Bogomolny2008, Bittner2010, Bogomolny2011, Hales2011, Bittner2011a, Bogomolny2012, Bittner2012a, Bittner2012d}. In the following two subsections we will introduce the trace formula for the dielectric square resonator, discuss its connection to the ray-based model for the dielectric square, and compare its predictions with the measured spectral data.

\subsection{The trace formula for the dielectric square resonator}
The DOS of an open cavity is given by
\begin{equation} \varrho(k) = -\frac{1}{\pi} \sum \limits_j \frac{\Im{k_j}}{[k - \Re{k_j}]^2 + [\Im{k_j}]^2} \end{equation}
where the $k_j$ are the resonance wave numbers \cite{Bogomolny2008}. It can be written as the sum of a smooth and a fluctuating part. The former is known as the Weyl term $\rhow$ and is the derivative $\rhow(k) = \dbyd{N}{k}$ of the smooth part of the resonance counting function $N(k)$, which for a 2D dielectric resonator with refractive index $n$ is given by
\begin{equation} N(k) = \frac{A n^2}{4 \pi} k^2 + \tilde{r}(n) \frac{L}{4 \pi} k \, . \end{equation}
Here $A = a^2$ is the area and $L = 4 a$ is the circumference of the resonator and
\begin{equation} \tilde{r}(n) = \frac{4 n}{\pi} \mathrm{E}\left[ \frac{n^2-1}{n^2} \right] - n \end{equation}
with $\mathrm{E}(x)$ the complete elliptic integral of the second kind \cite{Abramowitz1972}. In the case discussed here, $n$ is the effective refractive index $\neff$ and hence exhibits a nonnegligible dispersion. Accordingly, $n$ will be treated as a frequency dependent quantity in the following. In the semiclassical limit $k \rightarrow \infty$ the fluctuating part of the DOS, $\rhof$, can be expressed as a sum over the periodic orbits of the corresponding classical (billiard) system. In the case of the dielectric square resonator it is given by
\begin{equation} \label{eq:trForm} \begin{array}{rcl} \rhofscl(k) & = & \sqrt{\frac{k}{2 \pi^3}} \sum \limits_{n_x = 1}^\infty \sum \limits_{n_y = 0}^{n_x} F_{n_x, n_y} \sqrt{n} \left( n + k \dbyd{n}{k} \right) \frac{a^2}{\sqrt{\lpo(n_x, n_y)}} \\ & & \times [r(\chi_x)]^{2 n_x} [r(\chi_y)]^{2 n_y} e^{i[ k n \lpo(n_x, n_y) - \pi/4 ]} + \mathrm{c.c.} \end{array} \end{equation}

\begin{figure}[tb]
\begin{center}
\includegraphics[width = 8.0 cm]{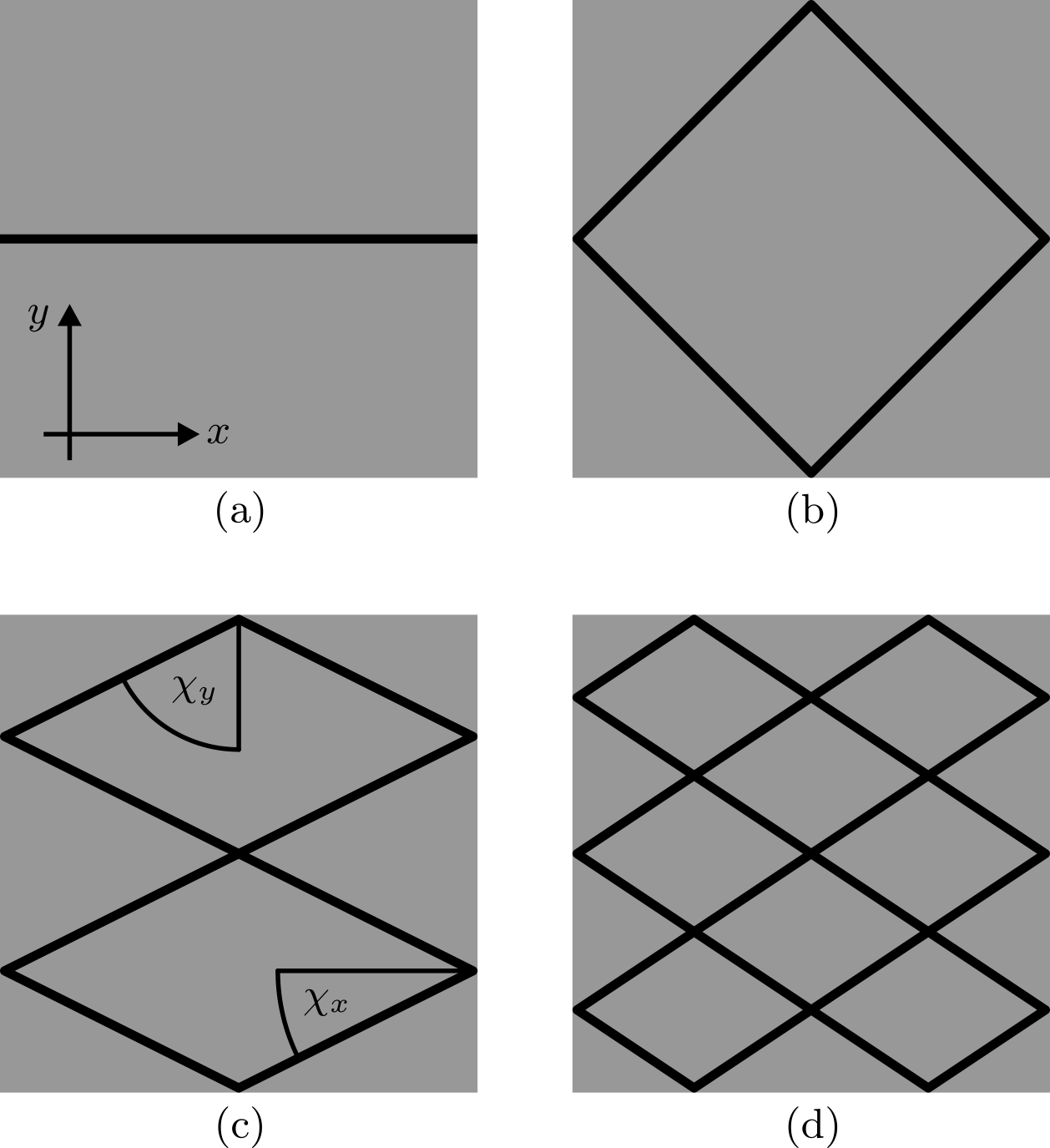}
\end{center}
\caption{Examples of POs in the square billiard. (a) $(1, 0)$ orbit, which is also called Fabry-P\'erot orbit, (b) $(1, 1)$ or diamond orbit, (c) $(2, 1)$ orbit, and (d) $(3, 2)$ orbit. The angles of incidence of an orbit on the edges perpendicular to the $x$ ($y$) axis are $\chi_x$ ($\chi_y$).}
\label{fig:POs}
\end{figure}

This formula can be derived from the quantization condition, \refeq{eq:quantCond}, as detailed in Appendix \ref{sec:traceForm}. The indices $n_{x,y}$ denominate a family of POs of the square billiard, with $n_x$ ($n_y$) being half the number of reflections of the orbits at the side walls perpendicular to the $x$ ($y$) axis. Some examples are shown in \reffig{fig:POs}. The lengths of the POs are
\begin{equation} \lpo(n_x, n_y) = 2 a \sqrt{n_x^2 + n_y^2} \end{equation}
and the factor $F_{n_x, n_y}$ equals $F_{n_x, n_y} = 2$ if either $n_x = n_y$, $n_x = 0$, or $n_y = 0$, and $F_{n_x, n_y} = 4$ otherwise. The angles of incidence of the POs on the edges (cf.~\reffig{fig:POs}) are given by
\begin{equation} \chi_{x, y} = \arctan(n_{y, x} / n_{x, y}) \, . \end{equation}
It should be noted that the trace formula (\ref{eq:trForm}) is identical to the formula given in Ref.~\cite{Bogomolny2008} for dielectric resonators with regular classical dynamics except for an additional factor $(1 + \frac{k}{n} \dbyd{n}{k})$ accounting for the dispersion of $n$. The same factor was found in Ref.~\cite{Bittner2012a}.

\subsection{Comparison with the experimental length spectrum}

\begin{figure*}[tb]
\begin{center}
\includegraphics[width = 16.0 cm]{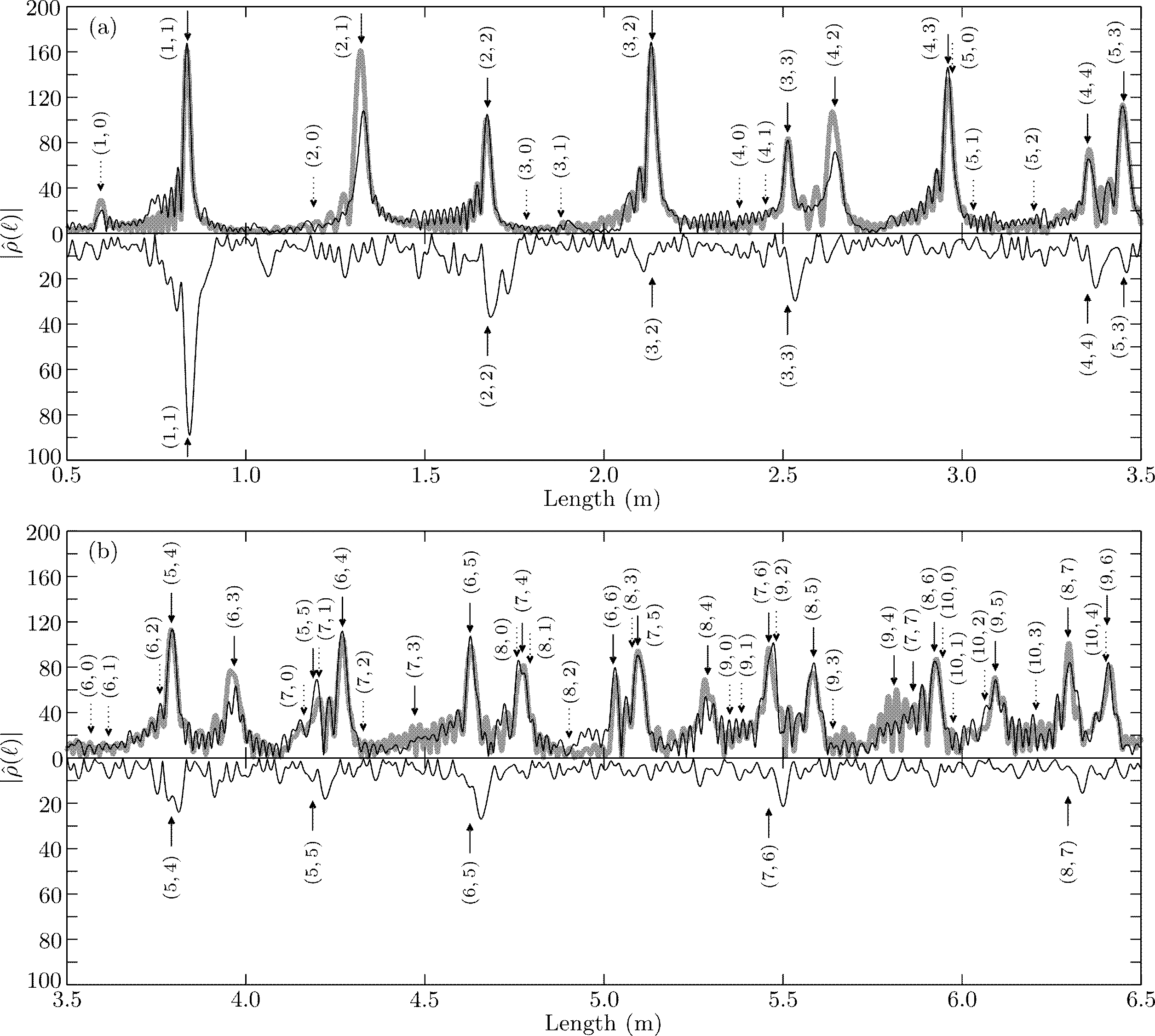}
\end{center}
\caption{Length spectrum $|\rhot(\ell)|$ and FT of the trace formula for the dielectric square resonator in the length regimes (a) $\ell = 0.5$--$3.5$~m and (b) $\ell = 3.5$--$6.5$~m. Each panel is divided into two parts. The solid line in the lower parts is the experimental length spectrum. In the upper parts, the gray line is the FT of the trace formula, \refeq{eq:trForm}, and the solid line is the length spectrum deduced from the set of modes that are confined by TIR. Note the different scales of the top and bottom parts. The arrows indicate the lengths $\lpeak \approx \lpo$ according to \refeq{eq:lpeak} of the POs labeled by their indices $(n_x, n_y)$. The dotted arrows denote those POs not confined by TIR. Only the POs that are clearly visible in the experimental length spectrum are indicated in the bottom parts.}
\label{fig:lenSpec}
\end{figure*}

We compared the FT of the measured DOS with the FT of the trace formula, \refeq{eq:trForm}. Here a modified definition of the FT was used to account for the dispersion of the refractive index $n$, 
\begin{equation} \label{eq:lspect} \begin{array}{rcl} \rhot(\ell) & = & \int \limits_{\kmin}^{\kmax} dk \, \rhof(k) \exp[-i k n(k) \ell] \\ \\ & = & \sum \limits_j e^{-i n(k_j) k_j \ell} - \mathrm{FT}\{ \rhow(k) \} \, , \end{array} \end{equation}
where $k_\mathrm{min, max} = 2 \pi f_\mathrm{min, max} / c$ correspond to the lower and upper bounds of the considered frequency range, respectively, and $\ell$ is the geometric length \cite{Bittner2012a}. Accordingly, the quantity $|\rhot(\ell)|$ is called the length spectrum. The experimental length spectrum $|\rhot_\mathrm{expt}(\ell)|$ is shown in the bottom parts of \reffig{fig:lenSpec}. It should be noted that the $(-+)$ modes were counted doubly due to their degeneracy with the $(+-)$ modes, yielding a total of $222$ resonances in the considered range of $\fmin = 5.5$~GHz to $\fmax = 10.0$~GHz (cf.\ \reftab{tab:WFsymVSant}). There are several peaks that stick out of the background noise, which has an average amplitude of about $\left< |\rhot(\ell)| \right> = 10$. The positions of these agree approximately with the lengths of the different POs and are thus related to these POs as predicted by the trace formula. The POs are indicated by their indices $(n_x, n_y)$, and the arrows indicate the expected peak positions. These deviate somewhat from the geometric lengths of the POs due to the dispersion of the refractive index on which the reflection phase shifts depend and can be estimated as \cite{Bittner2012a}
\begin{equation} \label{eq:lpeak} \begin{array}{rcl} \lpeak(n_x, n_y) & = & \lpo(n_x, n_y) + 2 n_x \left. \frac{\pbyp{\arg(r)}{n}(\chi_x) \dbyd{n}{k}}{n + k \dbyd{n}{k}} \right|_{k_0(\chi_x)} \\ & & + 2 n_y \left. \frac{\pbyp{\arg(r)}{n}(\chi_y) \dbyd{n}{k}}{n + k \dbyd{n}{k}} \right|_{k_0(\chi_y)} \, . \end{array} \end{equation}
Here, 
\begin{equation} k_0(\chi) = \left\{ \begin{array}{ccl} (\kmin + \kmax) / 2 & : & \fcrit(\chi) < \fmin \\ (\kcrit(\chi) + \kmax) / 2 & : & \fmin \leq \fcrit(\chi) \end{array} \right. \, , \end{equation}
where $\kcrit = 2 \pi \fcrit / c$ and the critical frequency is defined via $\sin(\chi) = 1 / n(\fcrit)$. The second and third terms in \refeq{eq:lpeak} therefore vanish when $\chi_x$ and $\chi_y$ are smaller than the critical angles $\alphacrit$ in the whole considered frequency range, respectively. This shift of the peak positions in the length spectrum with respect to the lengths of the POs should not be confused with the Goos-H{\"a}nchen shift. 

For comparison, the FT of the trace formula (called the semiclassical length spectrum in the following, $|\rhot_\mathrm{scl}(\ell)|$) is depicted as thick gray line in the upper parts of \reffig{fig:lenSpec}. The semiclassical length spectrum features a large number of peaks that almost all correspond to POs confined by TIR, i.e., for which $\fcrit(\chi_\po) < \fmax$, where $\chi_\po = \min\{ \chi_x, \chi_y \}$. These are indicated by solid arrows in \reffig{fig:lenSpec}, whereas those associated with POs that are not confined are indicated by dotted arrows. In the experimental length spectrum all the visible peaks are related to POs confined by TIR like in Refs.~\cite{Bittner2010, Bittner2012a, Bittner2012d}, though not all of these POs are visible in the experimental length spectrum. So while the experimental length spectrum shows good qualitative agreement with the trace formula prediction, there are quantitative deviations. First, the peak positions of the experimental length spectrum, $\lpeakexp$, are slightly but systematically shifted with respect to the peak positions of the semiclassical length spectrum, $\lpeakscl$. Second, the peak amplitudes $\Aexp = |\rhot_\mathrm{expt}(\lpeakexp)|$ are smaller than those of the semiclassical length spectrum, $\Ascl = |\rhot_\mathrm{scl}(\lpeakscl)|$. These findings are summarized in \reftab{tab:lspect}.

\begin{table*}[tb]
\caption{Summary of the POs observed in the experimental length spectrum. The first column indicates the indices $(n_x, n_y)$ of the POs, the second their angles of incidence $\chi_\po$, the third the corresponding critical frequency $\fcrit(\chi_\po)$, the fourth their lengths $\lpo$, the fifth the expected peak position $\lpeak$ according to \refeq{eq:lpeak}, the sixth and seventh the actual peak positions $\lpeakexp$ and $\lpeakscl$ in the experimental and semiclassical length spectrum, respectively, and the eights and ninths the corresponding peak amplitudes $\Aexp$ and $\Ascl$.}
\label{tab:lspect}
\begin{center}
\begin{tabular}{c|c|r|r|r|r|r|r|r}
\hline
\hline
$(n_x, n_y)$ & $\chi_\po$ & $\fcrit$ (GHz) & $\lpo$ (m) & $\lpeak$ (m) & $\lpeakexp$ (m) & $\lpeakscl$ (m) & $\Aexp$ & $\Ascl$ \\
\hline
$(1, 1)$ & $45.0^\circ$ & $5.367$ & $0.841$ & $0.838$ & $0.843$ & $0.835$ & $88.9$ & $161.2$ \\
$(2, 2)$ & $45.0^\circ$ & $5.367$ & $1.682$ & $1.675$ & $1.684$ & $1.672$ & $37.0$ & $100.3$ \\
$(3, 2)$ & $33.7^\circ$ & $6.376$ & $2.144$ & $2.135$ & $2.132$ & $2.133$ & $16.9$ & $162.3$ \\
$(3, 3)$ & $45.0^\circ$ & $5.367$ & $2.523$ & $2.513$ & $2.534$ & $2.514$ & $29.8$ & $83.4$ \\
$(4, 4)$ & $45.0^\circ$ & $5.367$ & $3.364$ & $3.350$ & $3.373$ & $3.355$ & $24.1$ & $74.0$ \\
$(5, 3)$ & $31.0^\circ$ & $6.774$ & $3.467$ & $3.452$ & $3.460$ & $3.449$ & $17.3$ & $113.2$ \\
$(5, 4)$ & $38.7^\circ$ & $5.858$ & $3.807$ & $3.792$ & $3.813$ & $3.792$ & $23.9$ & $112.7$ \\
$(5, 5)$ & $45.0^\circ$ & $5.367$ & $4.204$ & $4.188$ & $4.222$ & $4.205$ & $18.2$ & $52.5$ \\
$(6, 5)$ & $39.8^\circ$ & $5.760$ & $4.644$ & $4.626$ & $4.657$ & $4.629$ & $27.0$ & $97.9$ \\
$(7, 6)$ & $40.6^\circ$ & $5.695$ & $5.482$ & $5.461$ & $5.501$ & $5.460$ & $21.4$ & $96.6$ \\
$(8, 7)$ & $41.2^\circ$ & $5.649$ & $6.321$ & $6.296$ & $6.336$ & $6.298$ & $15.7$ & $100.9$ \\
\hline
\hline
\end{tabular}
\end{center}
\end{table*}

\begin{figure}[tb]
\begin{center}
\includegraphics[width = 8.0 cm]{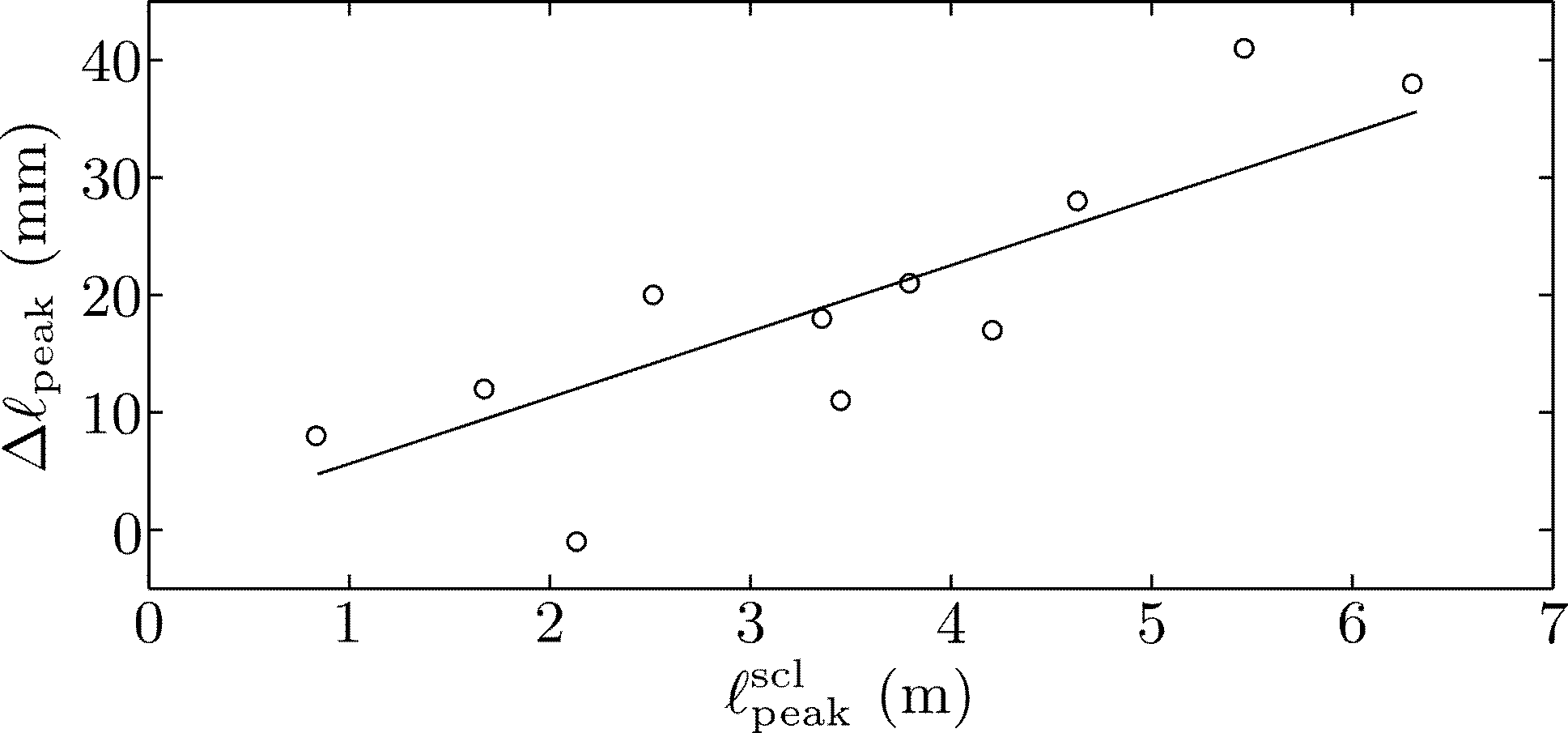}
\end{center}
\caption{Difference $\Delta \lpeak$ between the peak positions of the experimental and the semiclassical length spectrum versus the peak position $\lpeakscl$. The solid line is a linear fit.}
\label{fig:lpeakDev}
\end{figure}

The difference between the peak positions of the experimental and semiclassical length spectrum, $\Delta \lpeak = \lpeakexp - \lpeakscl$, are shown in \reffig{fig:lpeakDev}. They grow linearly with the length of the PO, and the solid line is a fit of the form $\Delta \lpeak = \mathcal{B} \lpeakscl$ with $\mathcal{B} = (5.63 \pm 2.70) \cdot 10^{-3}$. The same effect was observed in Ref.~\cite{Bittner2012a}, where it was attributed to the systematic deviations between the measured resonance frequencies and the predictions of the $\neff$ model. It was shown that $\mathcal{B}$ is equal to minus the derivative of $\Delta f = \fexp - \ftheo$ with respect to the frequency. This slope equals $B = (4.79 \pm 0.01) \cdot 10^{-3}$ [see \reffig{fig:freqComp} and \refeq{eq:freqDevFit}], which is in good agreement with $\mathcal{B}$. From this we can conclude that the frequency deviations evidenced in \reffig{fig:freqComp} are mainly due to the inaccuracy of the $\neff$ approximation and not due to that of the semiclassical model for the dielectric square.

\begin{figure}[tb]
\begin{center}
\includegraphics[width = 8.0 cm]{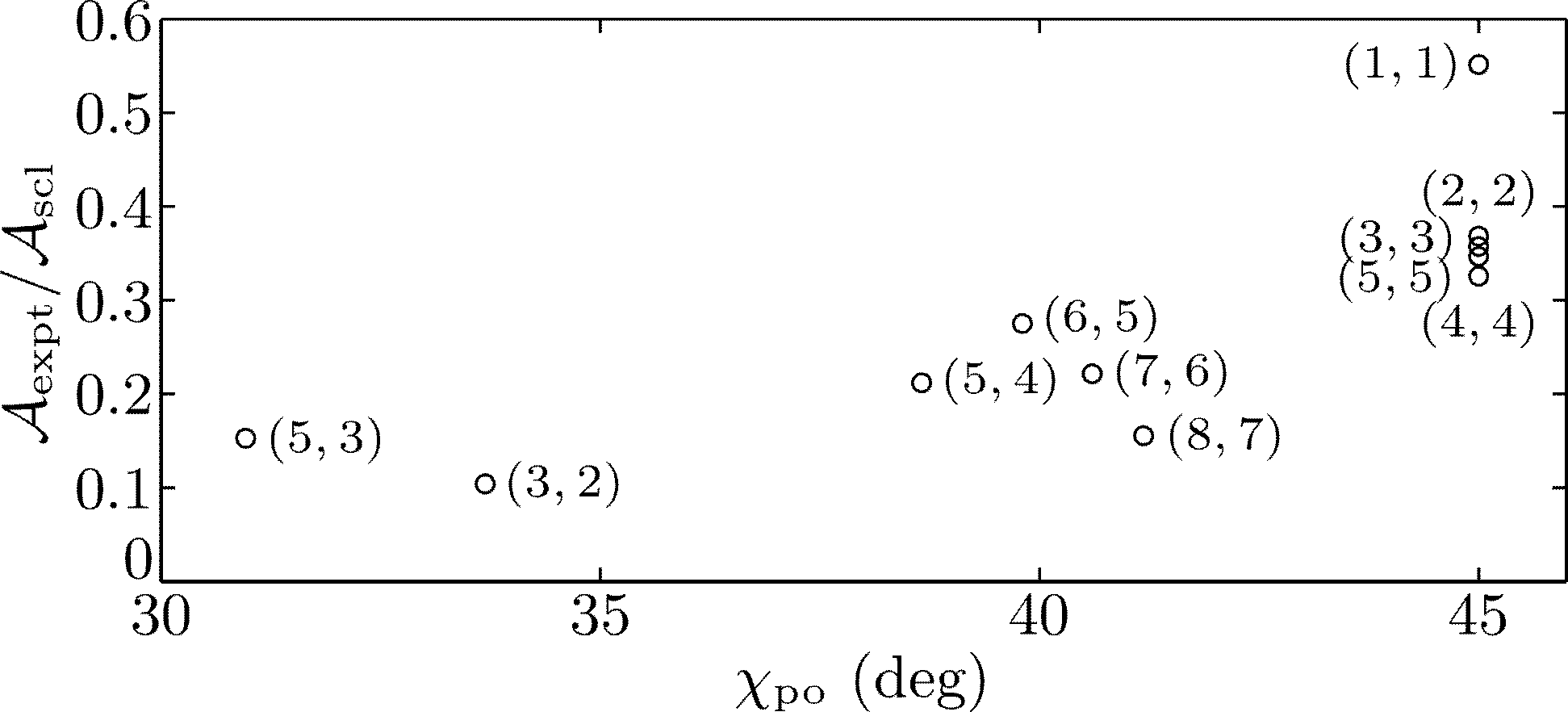}
\end{center}
\caption{Relative peak amplitudes $\Aexp / \Ascl$ for the POs observed in the experimental length spectrum versus their angle of incidence $\chi_\po$. The indices of the POs are indicated as $(n_x, n_y)$.}
\label{fig:relAmps}
\end{figure}

The ratios of the peak amplitudes of the experimental and semiclassical length spectrum, $\Aexp / \Ascl$, are depicted in \reffig{fig:relAmps} with respect to the angle of incidence of the POs, $\chi_\po$. They are in the range of $10$--$55\%$ and slowly decline with diminishing $\chi_\po$ (cf. Ref.~\cite{Bittner2010}) and their average value is $\left< \Aexp / \Ascl \right> = 27.9\%$. Given that the semiclassical model and the Weyl formula predict a total of about $1866$ modes in the given frequency range, so only about $11.9\%$ of all modes are actually observed experimentally, it is not surprising that the experimental peak amplitudes are significantly smaller. Since, however, all observed modes are related to trajectories with an angle of incidence above the critical angle (see \reffig{fig:alphaVsFreq}), it is interesting to compare the experimental length spectrum with the length spectrum for the set of calculated modes that are confined by TIR. It consists of $756$ modes, i.e., $40.5\%$ of all modes predicted by the model in the considered frequency range. The corresponding length spectrum is depicted as solid black line in the upper parts of \reffig{fig:lenSpec}. It agrees well with the semiclassical length spectrum except for a few peaks like that of the $(2, 1)$ orbit and its harmonics. This is related to the proximity of the $\chi_\po$ associated with that orbit to the critical angle and the inaccuracy of the stationary phase approximation for such orbits \cite{Bogomolny2008} and not to the lack of the nonconfined modes. In fact, the length spectrum obtained when including all $1866$ model modes (not shown) can hardly be distinguished from the one accounting only for the confined modes. This is expected because modes that are not confined contribute very little to the length spectrum since their finite imaginary part of $\ftheo$ leads to an exponential damping of their contribution [see \refeq{eq:lspect}]. So in practice, it suffices to consider only the confined modes to obtain the length spectrum predicted by the trace formula. Interestingly, the number of measured modes, $222$, is $29.4\%$ of the number of all confined modes, which is close to the mean ratio between experimental and semiclassical peak amplitudes in \reffig{fig:relAmps}. 

In summary we demonstrated that the trace formula for the dielectric square resonator is directly connected to our semiclassical model and that the indices $(n_x, n_y)$ of the POs are the conjugated variables of the quantum numbers $(m_x, m_y)$. Individual modes hence do not contribute to specific POs. On the other hand, it was shown in Refs.~\cite{Lebental2007, Bittner2010} that a spectrum containing only one or two families of modes related to trajectories with angle of incidence close to $\alphainc = 45^\circ$ leads to a length spectrum that exhibits only the diamond PO. The experimental length spectrum evidenced that the measured modes, that are all confined by TIR, correspond only to POs with this property in agreement with previous studies \cite{Bittner2010}. The qualitative agreement with the prediction deduced from the trace formula was good, and the deviations between the respective peak positions and amplitudes could be well explained with the inaccuracy of the $\neff$ model and the limited number of experimentally observed modes in accordance with Refs.~\cite{Bittner2010, Bittner2012a}.

\section{Conclusions} \label{sec:concl}
We investigated a dielectric square resonator in a microwave experiment with an alumina cavity. The spectra and field distributions were measured with various antenna positions on different symmetry axes of the square. The experimental data were compared with a simple semiclassical model \cite{Bittner2013b} and showed excellent agreement for an effective refractive index in the range of $\neff = 1.5$--$2.5$. The analysis of the momentum space representation of the field distributions proved particularly useful. First, it enabled a direct measurement of the effective refractive index of the resonator and the refractive index of the alumina. The measured values of $\neff$ furthermore validated the effective refractive index calculations with high precision. Second, contributions from different waveguide modes, i.e., $z$ excitations in the resonator, could be easily identified and removed within the momentum space representation. This allowed us to obtain field distributions with high data quality in frequency regimes that would be otherwise inaccessible experimentally. It should be noted that these two aspects apply in general to experiments with flat dielectric microwave resonators. Third, the association of the modes with classical tori in the square resonator was particularly evident in momentum space. It is presumed that similar phenomena exist also in other dielectric resonators with \mbox{(pseudo-)} integrable classical dynamics \cite{Lebental2007, Song2013a}. The ray-based model permitted us to identify the measured resonant states and label them with quantum numbers. This in turn allowed for a better understanding of the structure of the measured spectra, and it was shown that the modes of the dielectric square are organized in a very regular way akin to that of integrable systems \cite{Peres1984} as also evidenced in Ref.~\cite{Lebental2007}. Only modes associated with trajectories confined by TIR were observed since the resonator was passive. Microlasers, on the other hand, can also exhibit other modes due to their gain.

Depending on the symmetry of the excited resonant states, the excitation antenna was placed such that modes of specific symmetry classes were not excited. We came to the result that modes belonging to nondegenerate symmetry classes exhibit their symmetry regardless of the position of the excitation antenna, whereas for twofold degenerate modes the symmetry of the measured WFs depended strongly on it and could be partially controlled. This demonstrates the particular sensitivity of highly symmetric resonant structures not only to perturbations of their geometry but also to the manner of excitation. While the procedures and devices used to pump microlasers differ from those employed to excite microwave resonators, there are nonetheless techniques to influence the excitation of certain lasing modes by changing, e.g., the shape of the pumped domain \cite{Hentschel2008, Bachelard2012} or the wavelength and polarization of the pump beam \cite{Gozhyk2012}. While the aforementioned methods do not rely on modes with specific symmetries, the use of degenerate mode pairs might enable particularly simple and effective schemes to control the behavior of a laser. Furthermore, the length spectrum was investigated and yielded good qualitative agreement with the trace formula prediction. In addition it was demonstrated that the trace formula for the dielectric square can be directly derived from the ray-based model. Future projects are the extension of the model to the far-field distributions and a refined model that can also correctly predict the lifetimes of modes confined by TIR.

\begin{acknowledgments}
This work was supported by the Deutsche Forschungsgemeinschaft (DFG) within the Collaborative Research Center 634.
\end{acknowledgments}

\appendix
\section{Derivation of the trace formula} \label{sec:traceForm}
Here we derive the trace formula for the dielectric square resonator, \refeq{eq:trForm}, starting from the approximate quantization condition \refeq{eq:quantCond}. By introducing the variable $E = k^2$ the DOS can be written as 
\begin{equation} \varrho(E) = \sum \limits_{m_x, m_y = 0}^\infty \delta(E - \Exy) \end{equation}
where $\Exy = (k_x^2 + k_y^2) / n^2$ and $k_{x, y}$ are the solutions of the quantization condition \refeq{eq:quantCondSol}. Hence the function
\begin{equation} \gxy(E) = n^2 E - k_x^2 - k_y^2 \end{equation}
vanishes for $E = \Exy$ and
\begin{equation} \varrho(E) = \sum \limits_{m_x, m_y = 0}^\infty \left| \dbyd{\gxy}{E} \right| \delta[\gxy(E)] \, . \end{equation}
The summation over $m_{x, y}$ is rewritten as
\begin{equation} \sum \limits_{m_x, m_y = 0}^\infty [\dots] = \frac{1}{4} \sum \limits_{m_x, m_y = -\infty}^\infty [\dots] + \, \mathrm{additional \, terms} \end{equation}
where the additional terms yield contributions due to grazing orbits at the cavity boundaries \cite{Sieber1995}. Such higher-order corrections are ignored in the following. The derivative of $\gxy$ is
\begin{equation} \dbyd{\gxy}{E} = n^2 + 2 n E \dbyd{n}{E} - 2 \left[ \dbyd{k_x}{r}  \left. \dbyd{r}{n} \right|_{\alpha_x} + \dbyd{k_y}{r}  \left. \dbyd{r}{n} \right|_{\alpha_y} \right] \dbyd{n}{E} \, . \end{equation}
The last term is essentially the derivative of the Fresnel reflection phase in the case of modes confined by TIR. Since the reflection phase does not depend strongly on $n$ we neglect this term in the following. We can now transform
\begin{equation} \varrho(E) \simeq \frac{1}{4} \sum \limits_{m_{x, y} = -\infty}^\infty \left( n^2 + 2 n E \dbyd{n}{E} \right) \delta[\gxy(E)] \end{equation}
by means of the Poisson resummation formula to 
\begin{equation} \begin{array}{rcl} \varrho(E) &  \simeq & \frac{1}{4} \sum \limits_{n_{x, y} = -\infty}^\infty \int \limits_{-\infty}^{\infty} dm_x \int \limits_{-\infty}^{\infty} dm_y \left( n^2 + 2 n E \dbyd{n}{E} \right) \\ & & \times \delta(n^2 E - k_x^2 - k_y^2) e^{2 \pi i (m_x n_x + m_y n_y)} \, . \end{array} \end{equation}
The conjugated variables $n_{x, y}$ turn out to be the indices of the POs in the square billiard. The quantum numbers $m_{x, y}$ in the exponential are replaced using \refeq{eq:quantCondSol} and, furthermore, $dm_x dm_y \approx dk_x dk_y a^2 / (\pi^2)$ is used where again the derivative of the Fresnel coefficients was neglected. This yields
\begin{widetext}
\begin{equation} \varrho(E) \simeq \frac{a^2}{4 \pi^2} \sum \limits_{n_{x, y} = -\infty}^\infty \int \limits_{-\infty}^{\infty} dk_x \int \limits_{-\infty}^{\infty} dk_y \left( n^2 + 2 n E \dbyd{n}{E} \right) \delta(n^2 E - k_x^2 - k_y^2) e^{2 i a (k_x n_x + k_y n_y)} \, [r(\alpha_x)]^{2 n_x} [r(\alpha_y)]^{2 n_y} \, . \end{equation}
The integral is calculated by introducing polar coordinates and applying the stationary phase approximation to the azimuthal part. Furthermore, we revert to $\rho(k) = 2k \, \rho(E)$. The saddle point for $n_x = n_y = 0$ gives the area term of the Weyl formula $\rhow(k)$. The remaining terms yield the fluctuating part of the DOS in the semiclassical limit, \refeq{eq:trForm}.
\end{widetext}

\end{document}